\documentstyle[aps,prl,12pt,graphicx,epsfig,psfrag]{revtex}
\begin{document}

\title{Stress distribution in static two dimensional granular 
model media in the absence of friction}
\author{S. Luding}
\address{Institute for Computer Applications 1,
Pfaffenwaldring 27, 70569 Stuttgart, GERMANY\\
e-mail: lui@ica1.uni-stuttgart.de}

\date{to appear in Phys. Rev. E}

\maketitle

\begin{abstract}
We present simulations of static model sandpiles in two dimensions (2D)
and focus on the stress distribution in such arrays made
of discrete particles. We use the
simplest possible model, i.e. spherical particles with a linear
spring and a linear dashpot active on contact and without any 
frictional forces.
Our model is able to reproduce several recent theoretical predictions.
For different boundary conditions we examine the contact 
network and the stresses in the array and at the 
bottom of the pile. In some cases we observe a dip, i.e.
the relative minimum in pressure, under the center of the pile.
We connect the dip to arching, and we relate arching to 
the structure of the contact network. Finally, we
find that small polydispersity is sufficient to 
cause a so called stress-network, i.e. strong fluctuations
in stress. From those data we determine the probability 
distribution for the vertical stress at the bottom and relate
it to theoretical and other numerical work.
\end{abstract}


\def\PSFIGO{
\psfig{figure=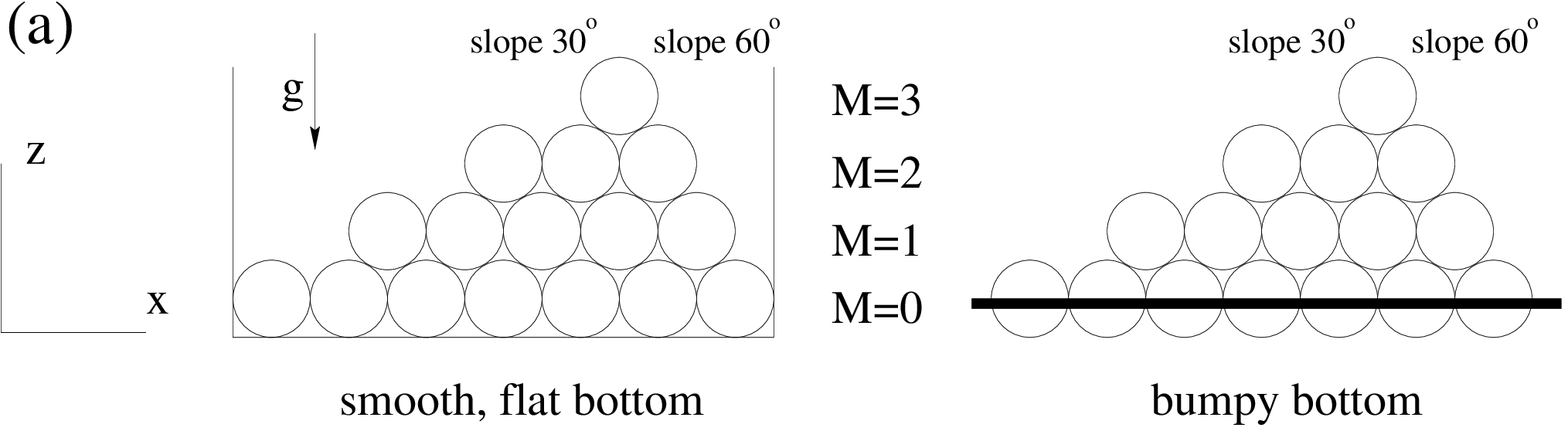,height=15.0cm,clip=,angle=-90}
\vskip 0.8cm
{~~~~~~~~~~~~~~~~~~~ \psfig{figure=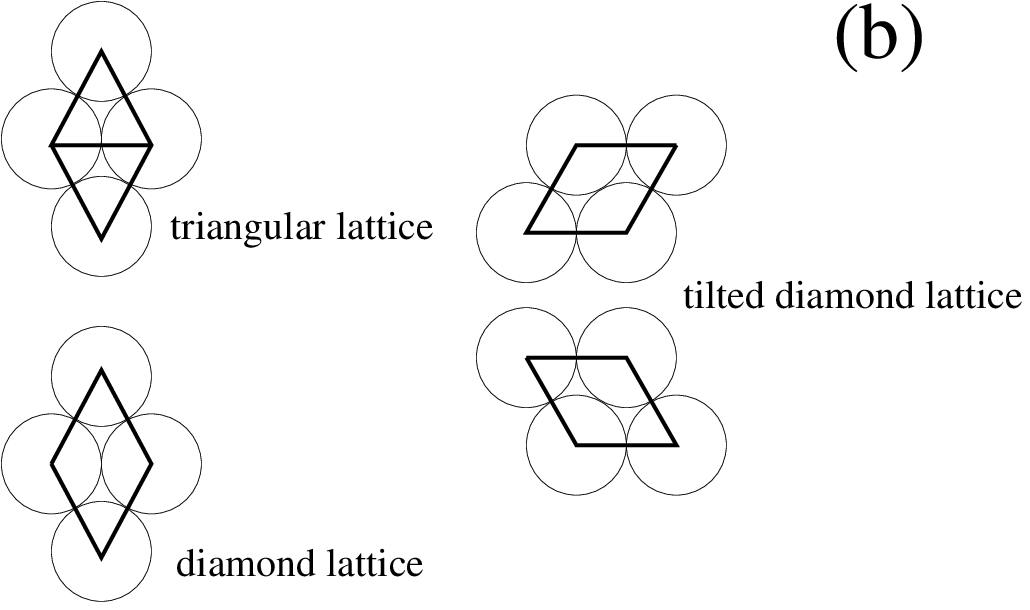,height=10.0cm,clip=,angle=-90}}
\begin{figure}[tbp]
\caption{
(a) Schematic drawing of a pile in a box with smooth, flat bottom (left),
and on a bumpy bottom (right), with $L_0 = 7$. 
The solid bar at the right indicates that the particles in row $M=0$ are 
fixed, so that the first relevant row with mobile particles is 
$M = 1$ with here $L_1 = 5$.
(b) Schematic drawing of the typical contact network configurations
in a regular arrangement.
}
\label{fig0}
\end{figure}
}

\def\PSFIGA{
\psfig{figure=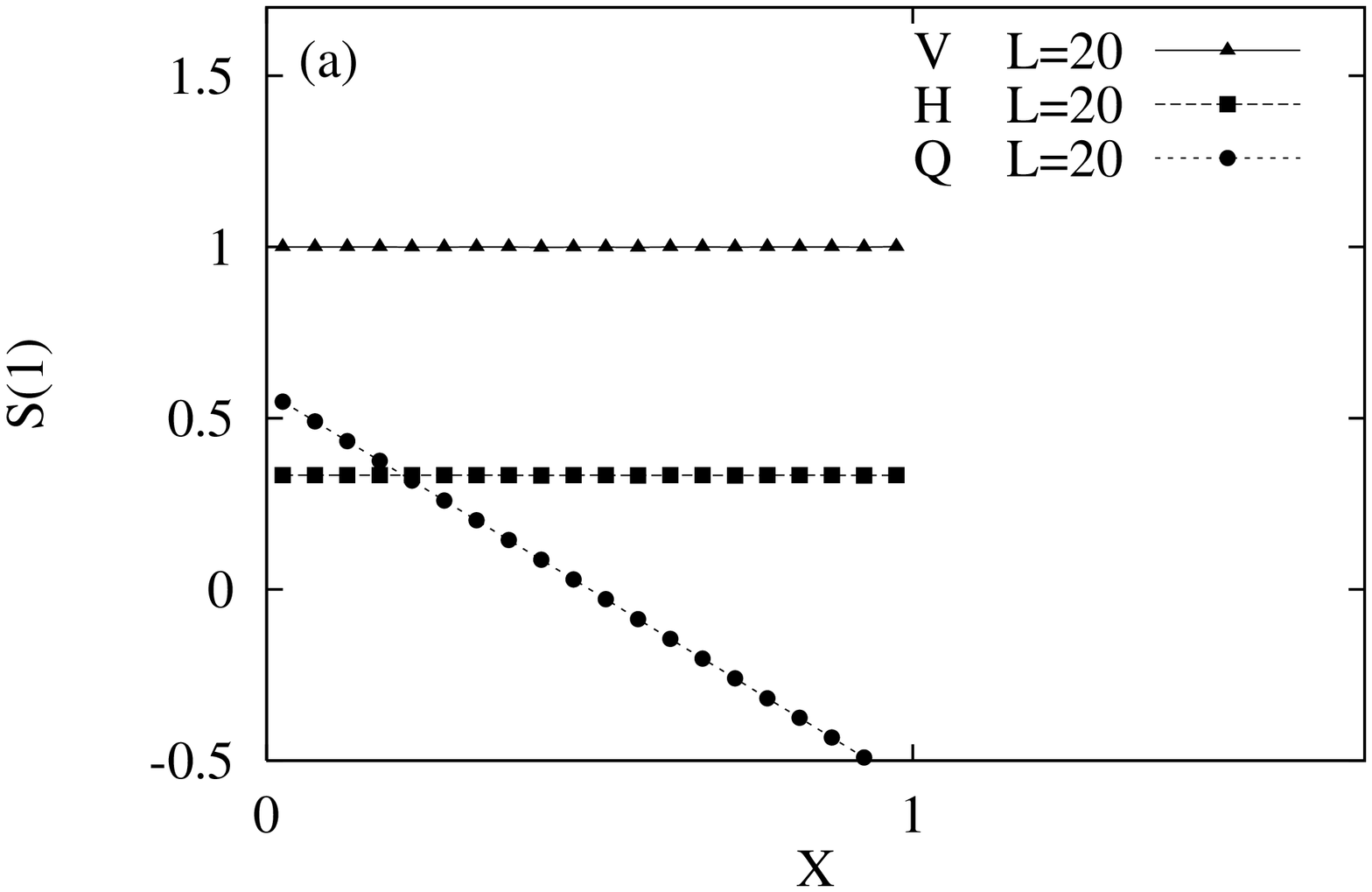,height=7.5cm,clip=,angle=-90}
\psfig{figure=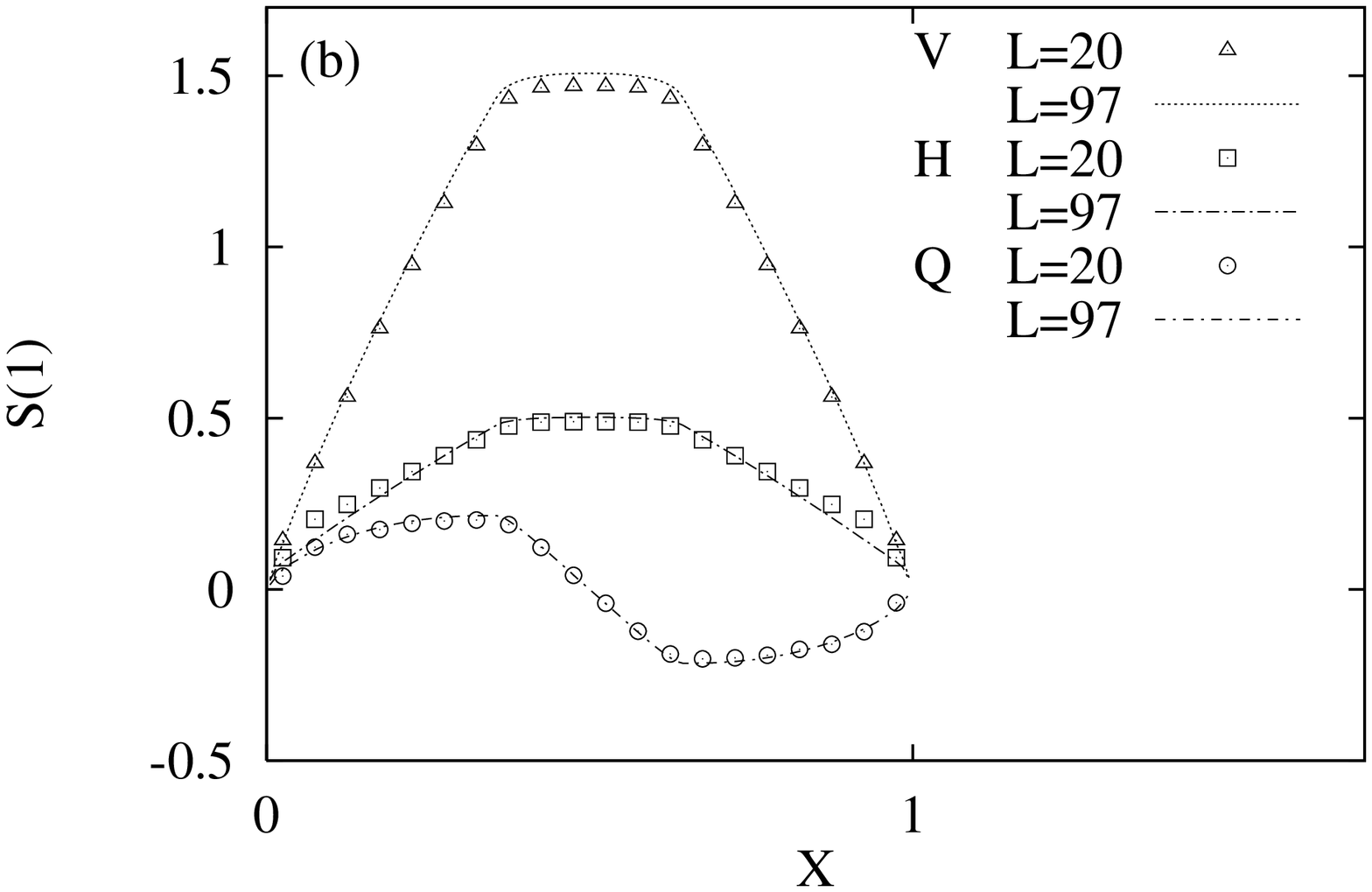,height=7.5cm,clip=,angle=-90}
\begin{figure}[tbp]
\caption{
Components of the dimensionless stress tensor $S(1)$ at row $M=1$ 
vs. dimensionless horizontal coordinate $X=x/l$, for a pile
with immobile particles at the bottom, $M=0$. The slope of the pile
is $60^o$ with $L_1=20$ in (a), and $30^o$ with $L_1=20$, or $L_1=97$ in (b).
We indicate the vertical stress with $V=S_{zz}$, the horizontal stress
with $H=S_{xx}$, and the shear stress with $Q=S_{xz}$.
}
\label{fig1}
\end{figure}
}
 
\def\PSFIGBA{
\begin{figure}[tbp]
\psfig{figure=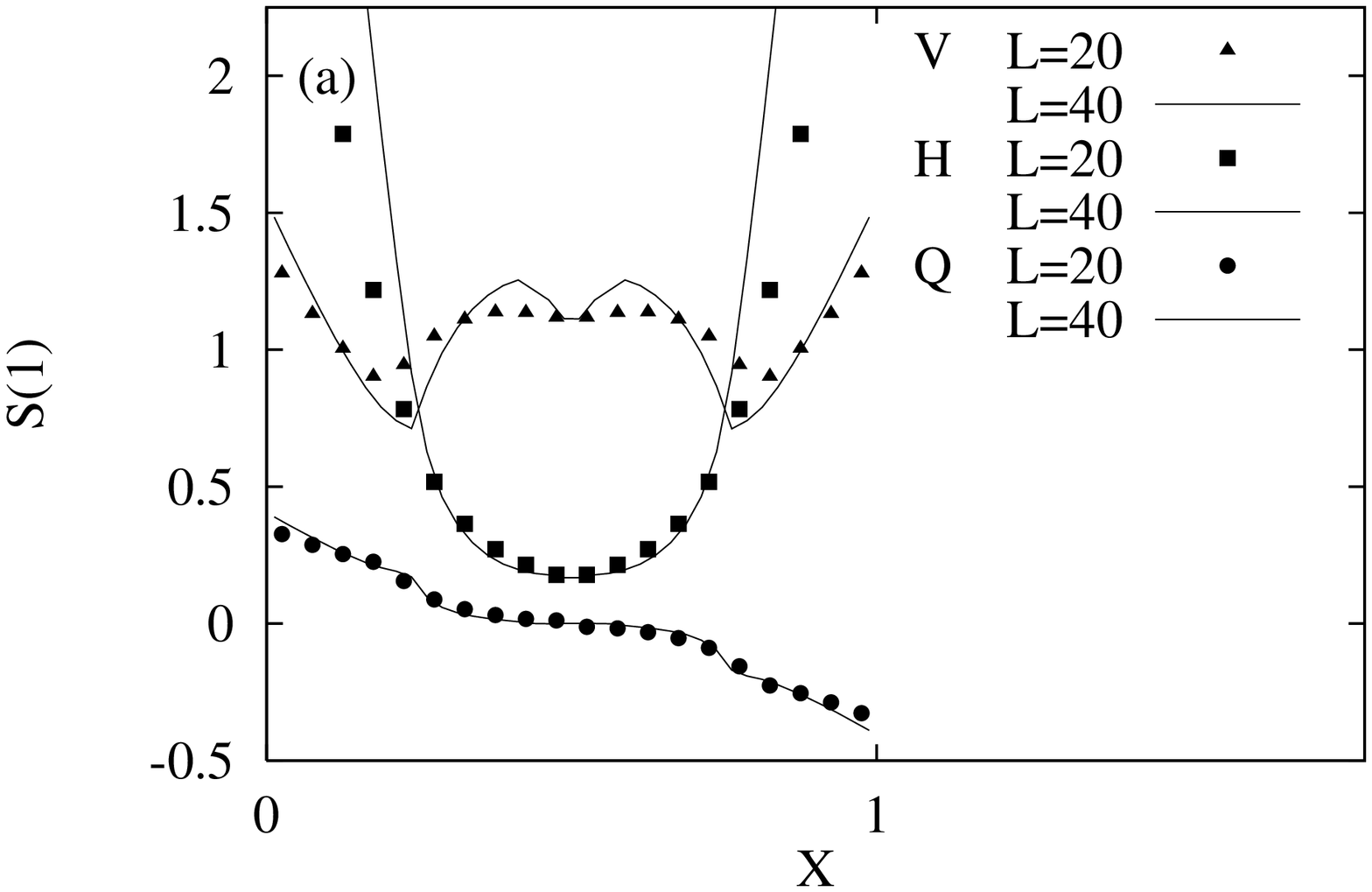,height=7.5cm,angle=-90,clip=}
\psfig{figure=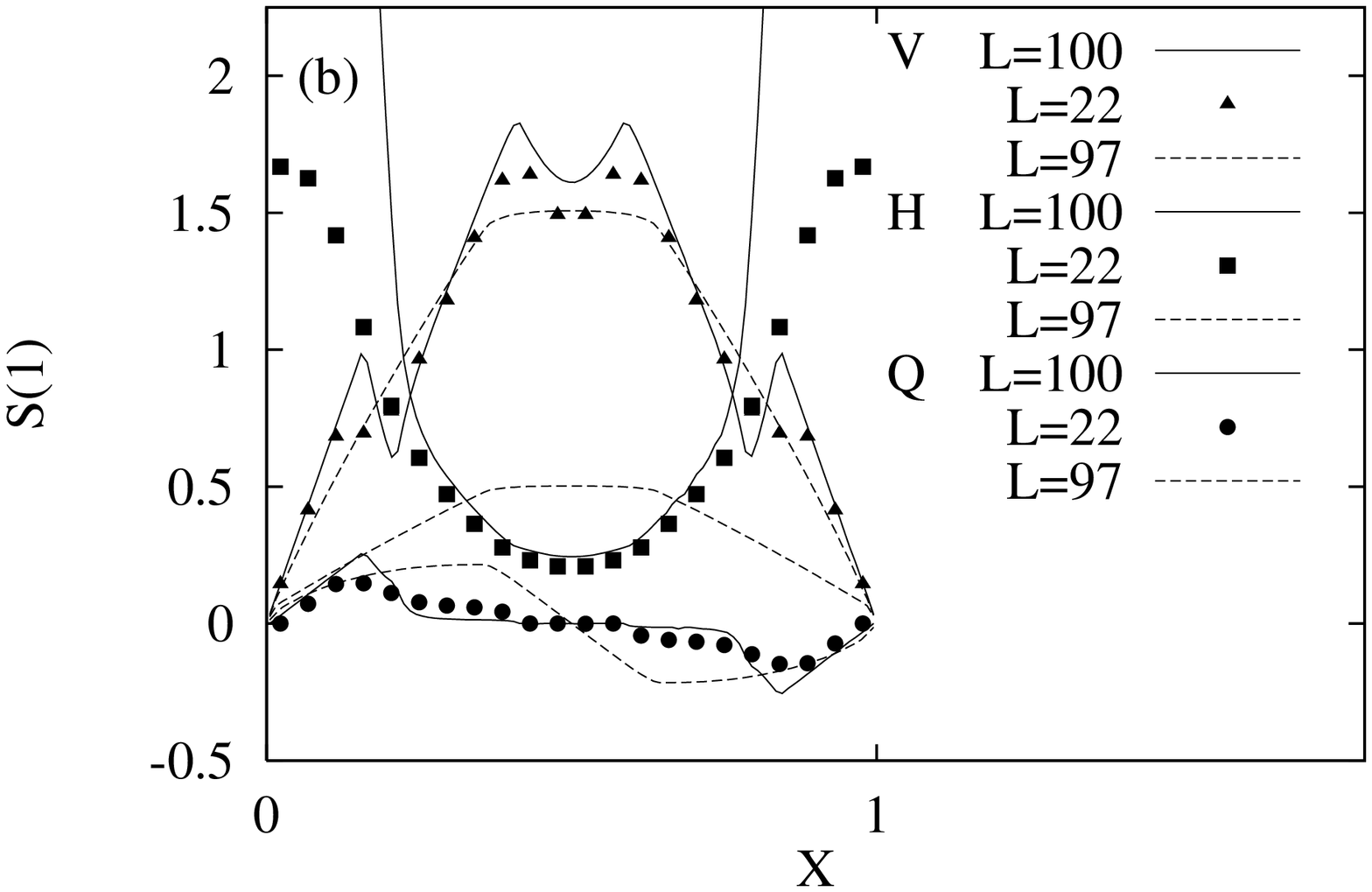,height=7.5cm,angle=-90,clip=}
\end{figure}
}
\def\PSFIGBB{
\begin{figure}[tbp]
\psfig{figure=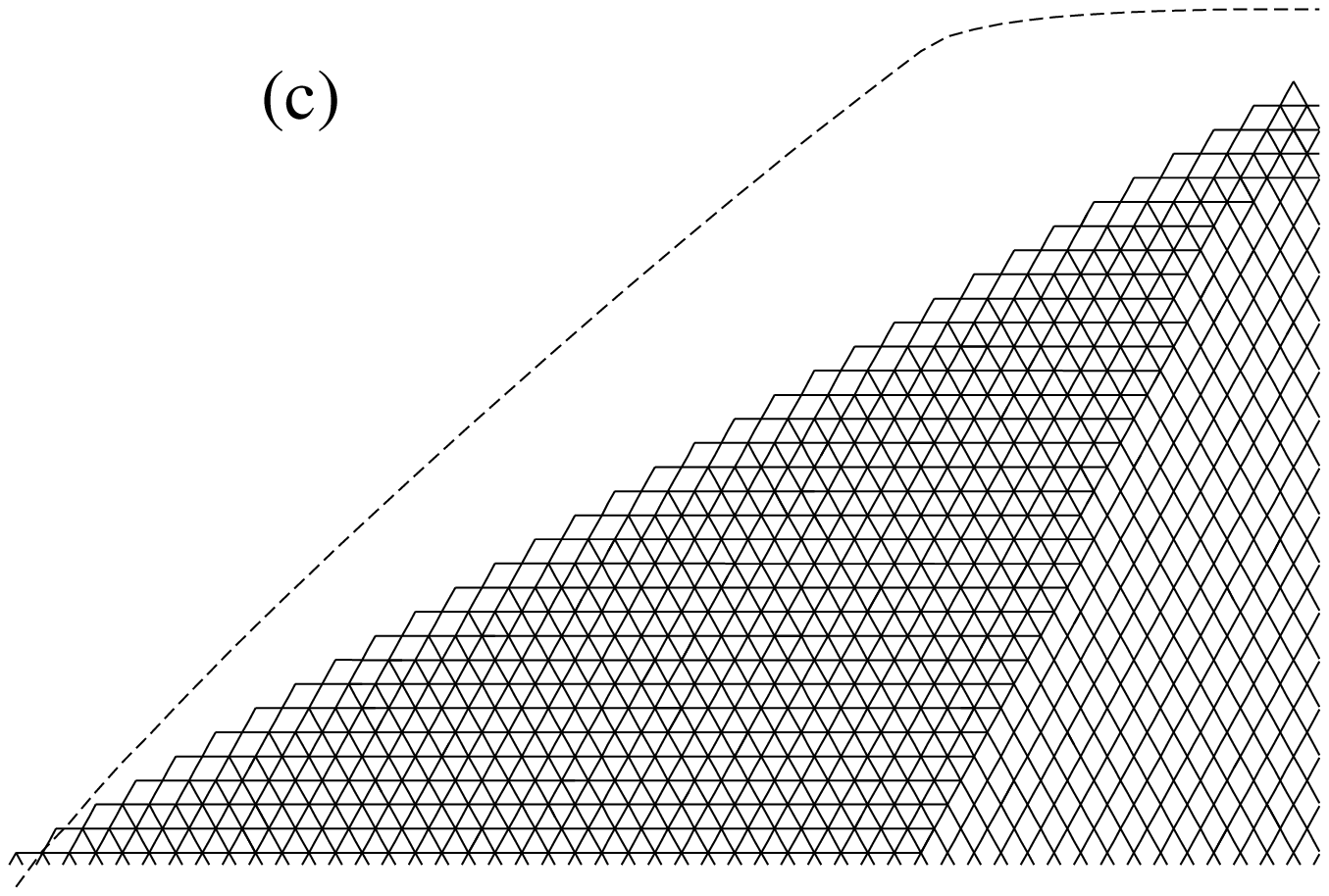,height=7.5cm,angle=-90,clip=}
\psfig{figure=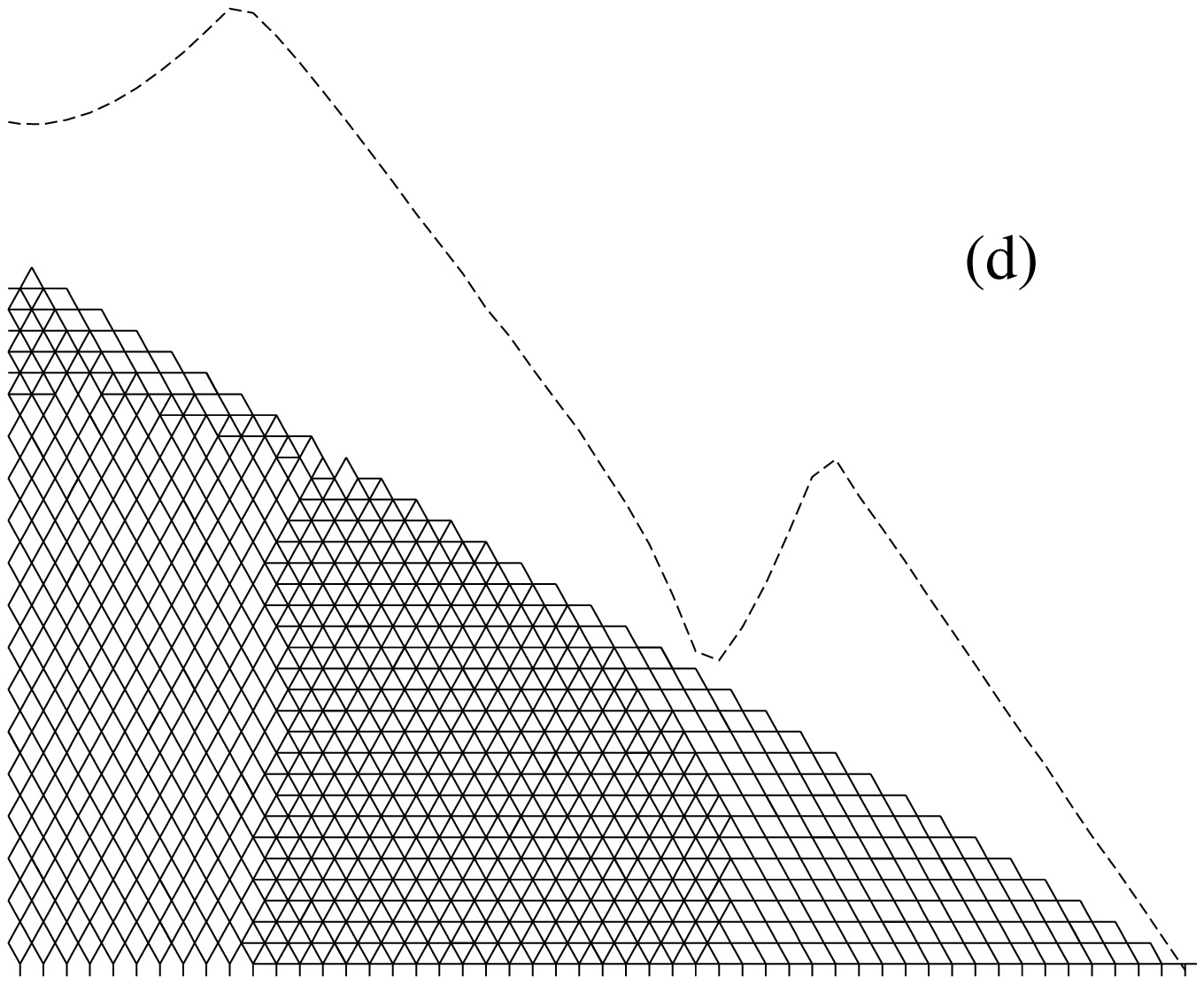,height=7.5cm,angle=-90,clip=}
\end{figure}
}
\def\PSFIGBC{
\begin{figure}[tbp]
\caption{
Components of the dimensionless stress tensor $S(0)$ vs. 
dimensionless horizontal coordinate $X=x/l$ at row $M=0$, for a pile
with mobile particles at the smooth and flat bottom. 
We indicate the vertical stress with $V=S_{zz}$, the horizontal stress
with $H=S_{xx}$, and the shear stress with $Q=S_{xz}$.
(a) The slope of the pile is $60^o$, and $L_0=20$ or $L_0=40$.
(b) The results for $30^o$ and $L_0=100$ (solid line) or $L_0=22$ 
    (symbols) are compared to the result for $L_1=97$ from 
    Fig.\ \protect \ref{fig1}(b). 
(c) Contact network for the left half of a $30^o$ pile 
    with $L_1=97$ and bumpy bottom. 
(d) Contact network for the right half of a $30^o$ pile 
    with $L_0=100$ and smooth, flat bottom. 
The dashed line in (c) and (d) gives the vertical stress $V$
for the corresponding piles.
}
\label{fig2}
\end{figure}
}

\def\PSFIGB{
\PSFIGBA
\PSFIGBB
\PSFIGBC
}

\def\PSFIGTA{
\begin{figure}[tbp]
\psfig{figure=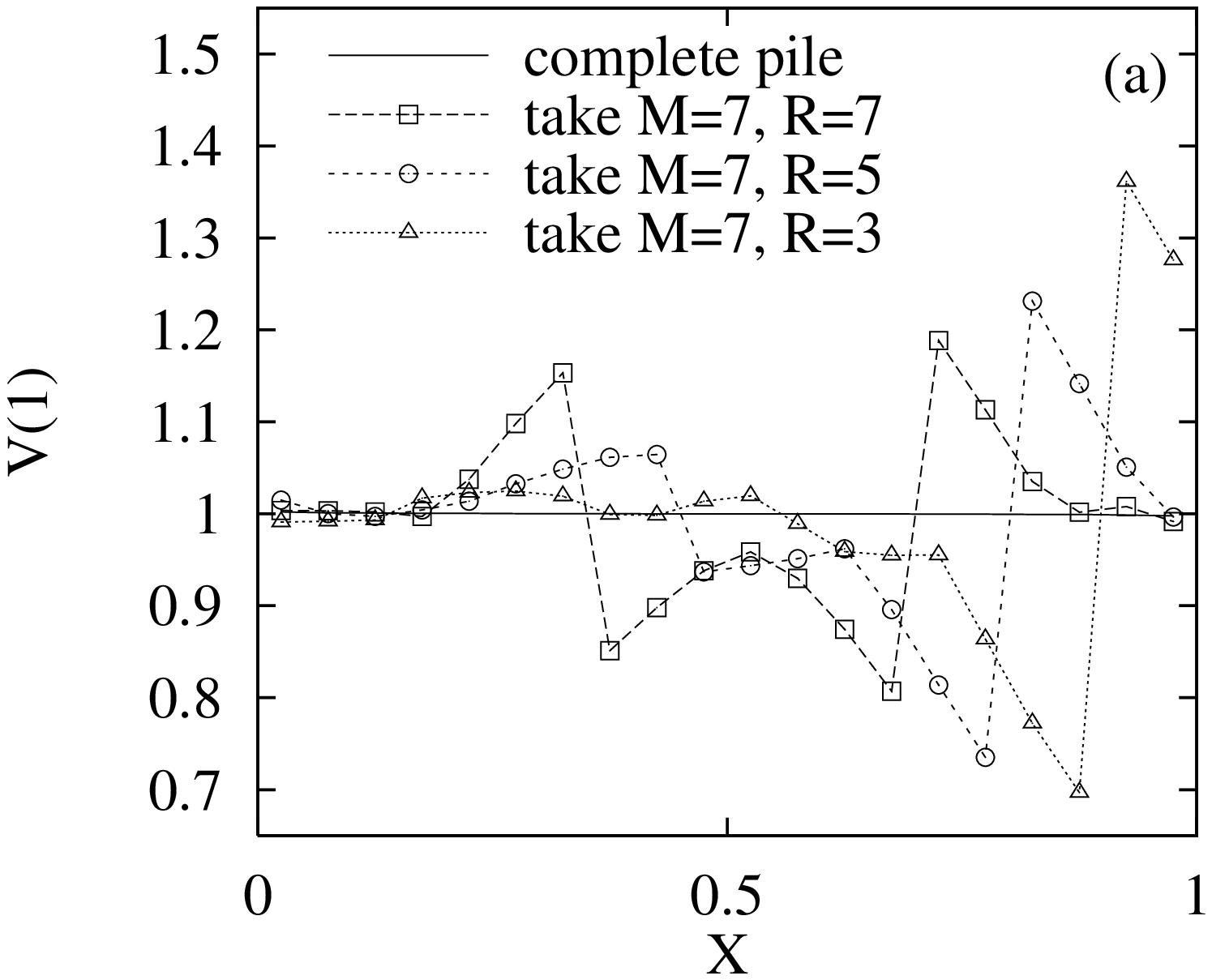,height=7.5cm,angle=-90,clip=}
\psfig{figure=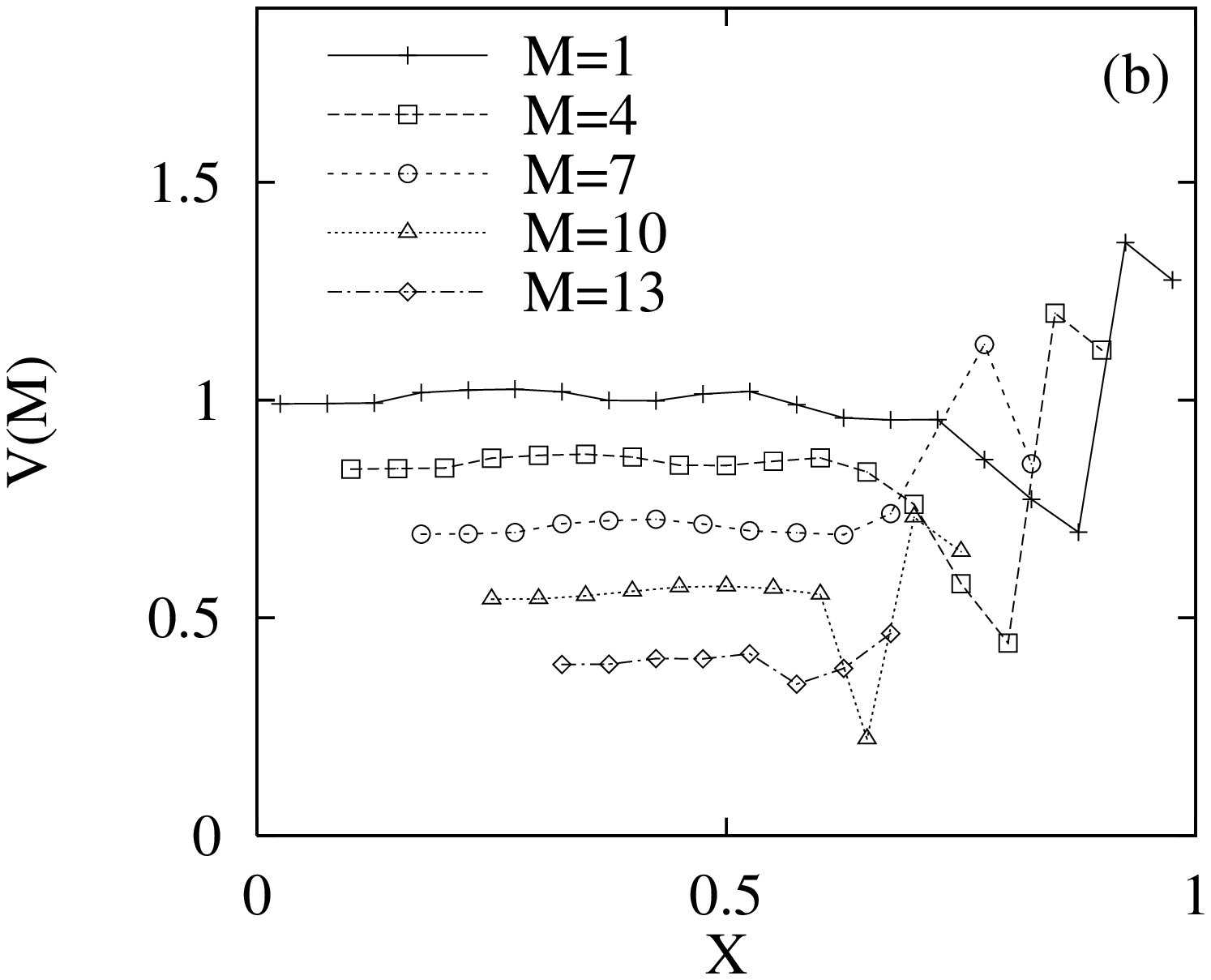,height=7.5cm,angle=-90,clip=}
\end{figure}
}
\def\PSFIGTB{
\begin{figure}[tbp]
\psfig{figure=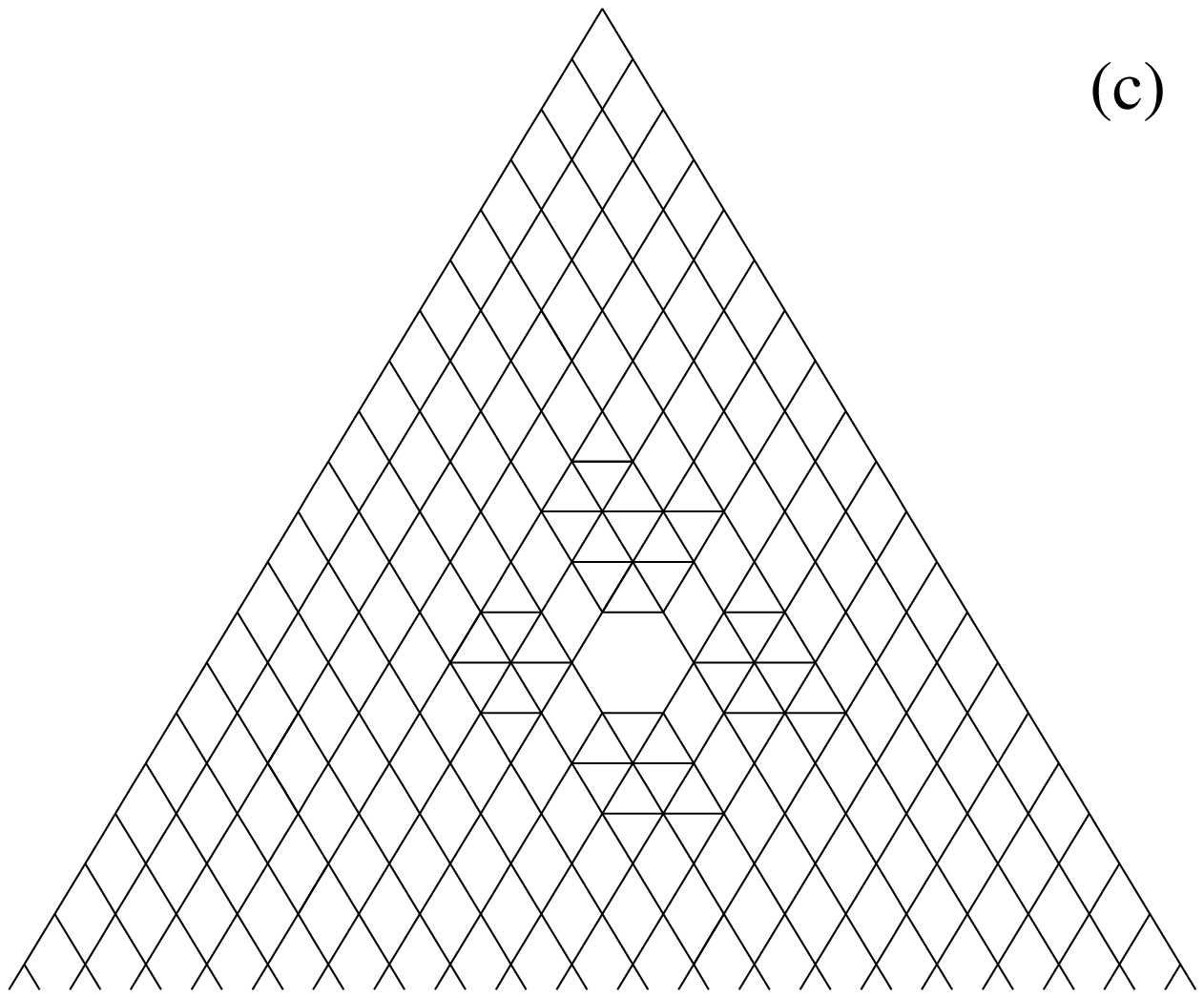,height=7.5cm,angle=-90,clip=}
\psfig{figure=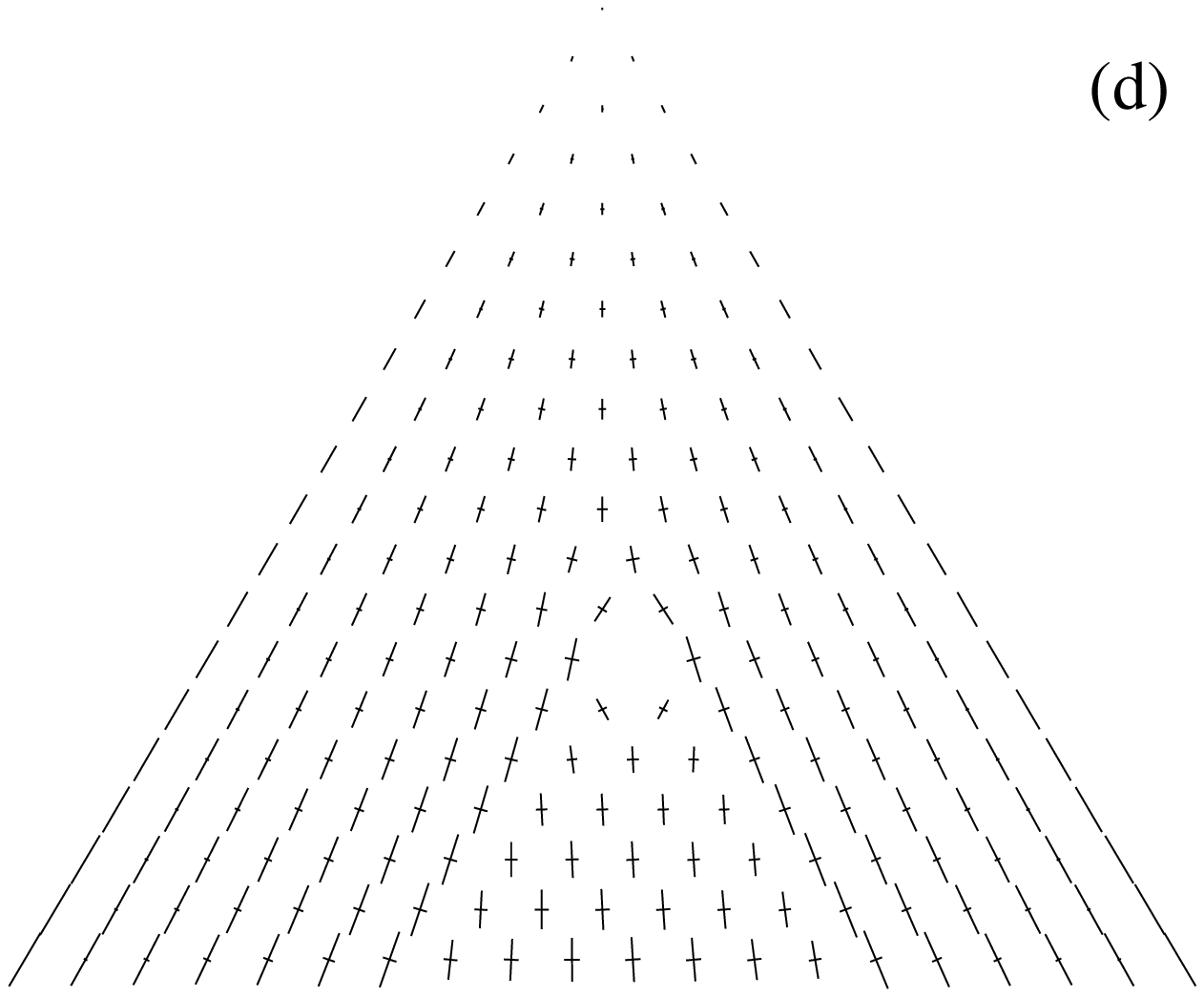,height=7.5cm,angle=-90,clip=}
\end{figure}
}
\def\PSFIGTC{
\begin{figure}[tbp]
\caption{
(a) Vertical stress $V(1)$ in row $M=1$ vs. $X$ for a $60^o$
    pile with bumpy bottom and $L_1=20$ (solid line). 
    $V(1)$ is given for piles where particle $R$ = 7, 5, or 3, 
    is removed from row 7; here $R$ counts from the right.
(b) The vertical stress $V(M)$ is plotted for different
    rows $M$ = 1, 4, 7, 10, 13, for the pile where particle 3
    is removed from row 7. Note the missing symbol for $M=7$
    (circles).
(c) Contact network for the situation where particle $R=7$ is
    removed from row $M=7$.
(d) Principal axis of stress for the situation where particle $R=7$ is
    removed from row $M=7$.
}
\label{figt}
\end{figure}
}

\def\PSFIGT{
\PSFIGTA
\PSFIGTB
\PSFIGTC
}

\def\PSFIGC{
\begin{figure}[tbp]
\psfig{figure=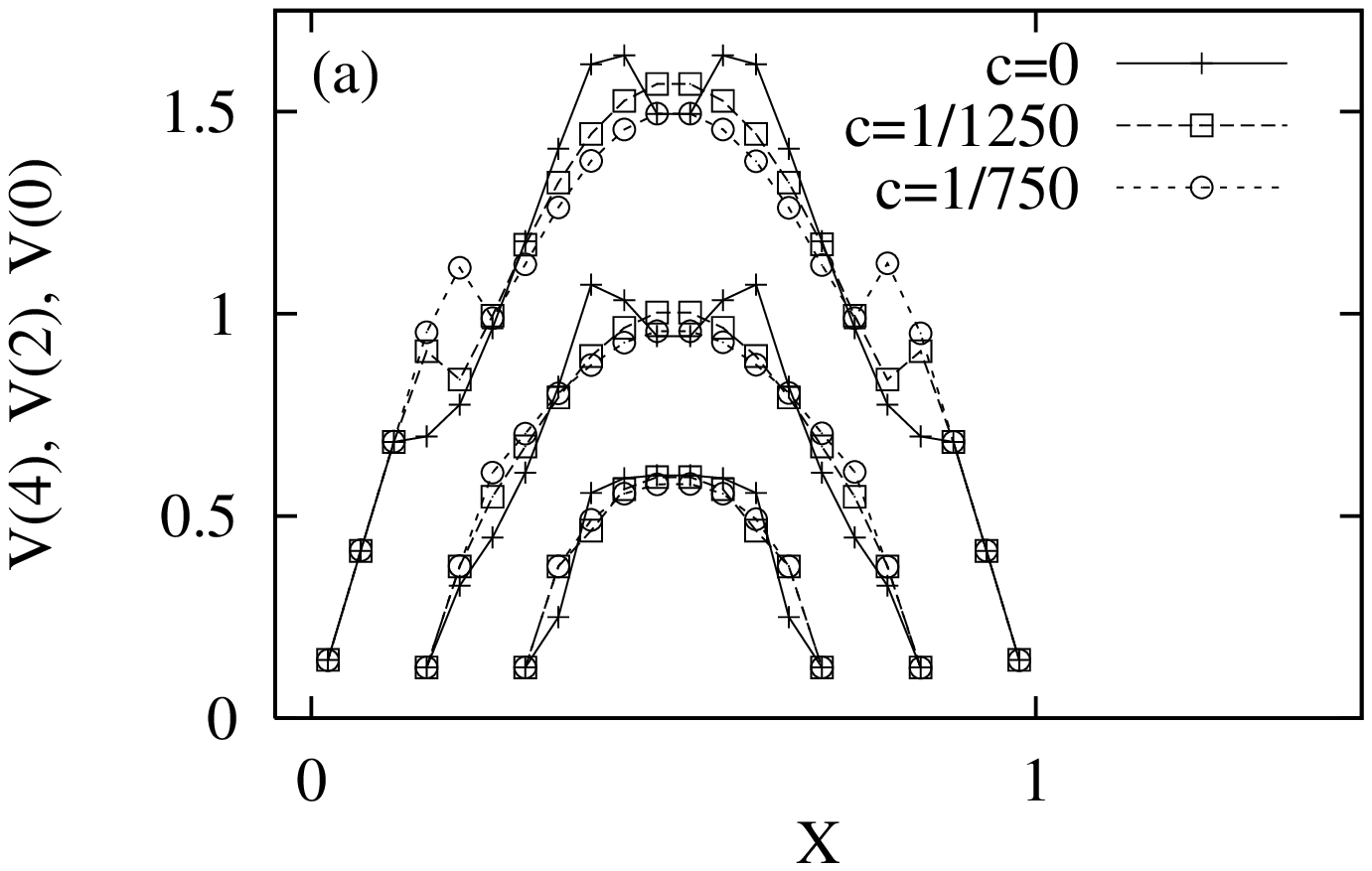,height=7.5cm,angle=-90,clip=}
\psfig{figure=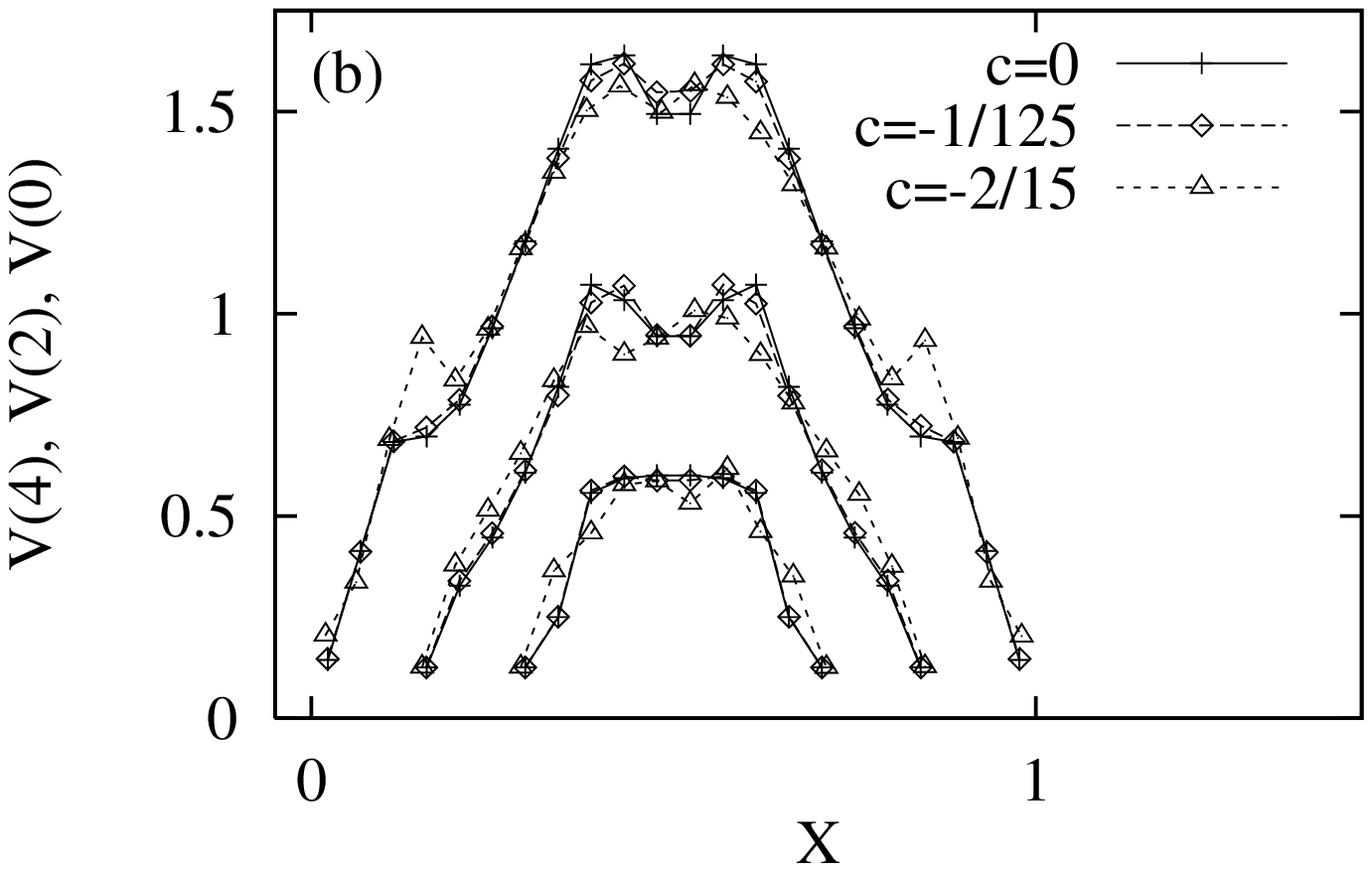,height=7.5cm,angle=-90,clip=}
\caption{
Vertical stresses $V(0), V(2), V(4)$,
in rows $M=0$, 2, and 4 respectively (from top to bottom) 
vs. $X$ for a $30^o$
pile with smooth, flat bottom and $L=22$. The two outermost particles
are fixed by vertical walls and have diameter $d=(1+c)d_0$.
The inserts gives the relative change $c$. (a) Large boundary particles
$c > 1$, and (b) small boundary particles $c < 1$.
}
\label{fig3}
\end{figure}
}

\def\PSFIGDA{
\begin{figure}[tbp]
\psfig{figure=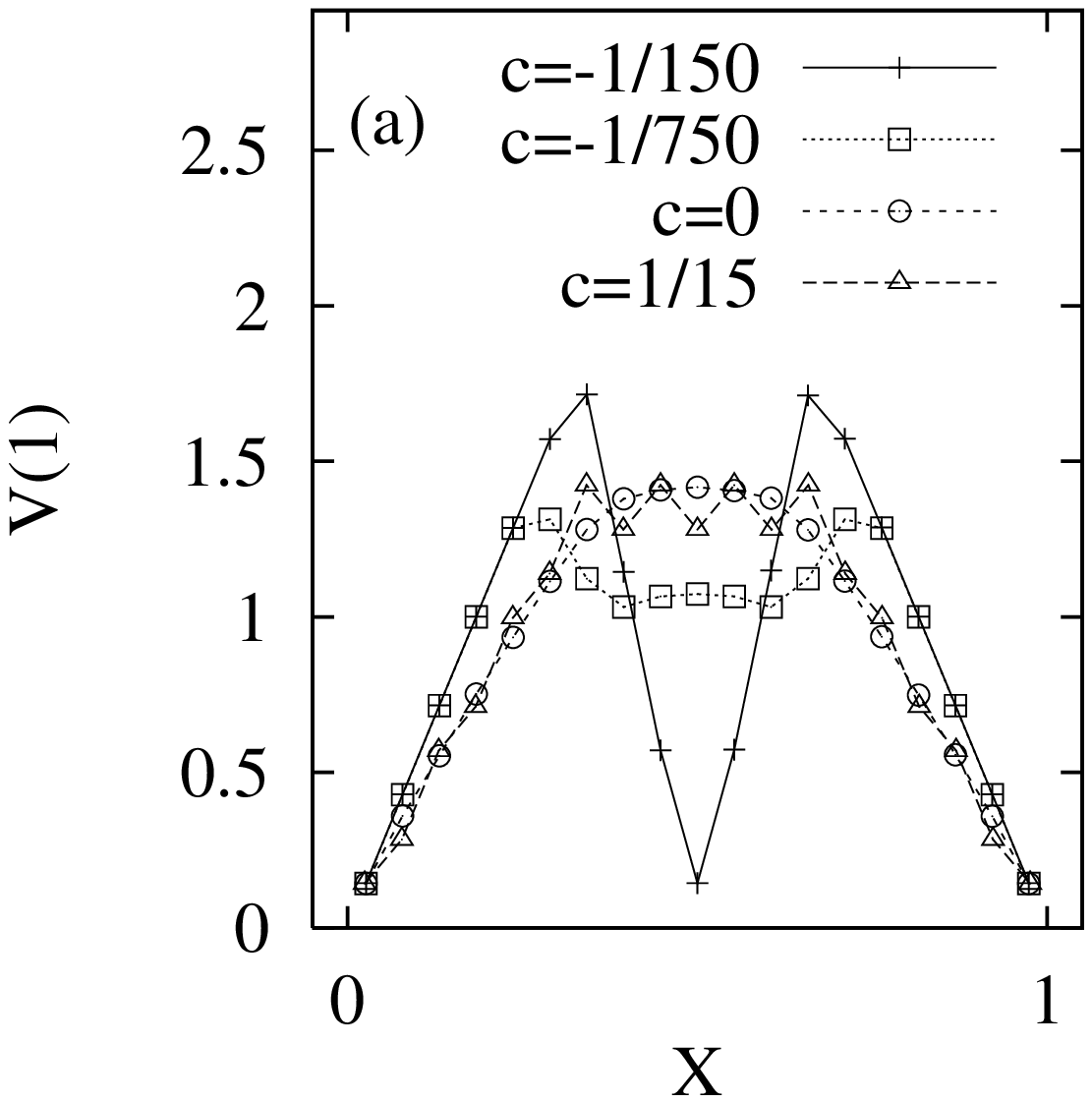,height=5.0cm,angle=-90,clip=}
\psfig{figure=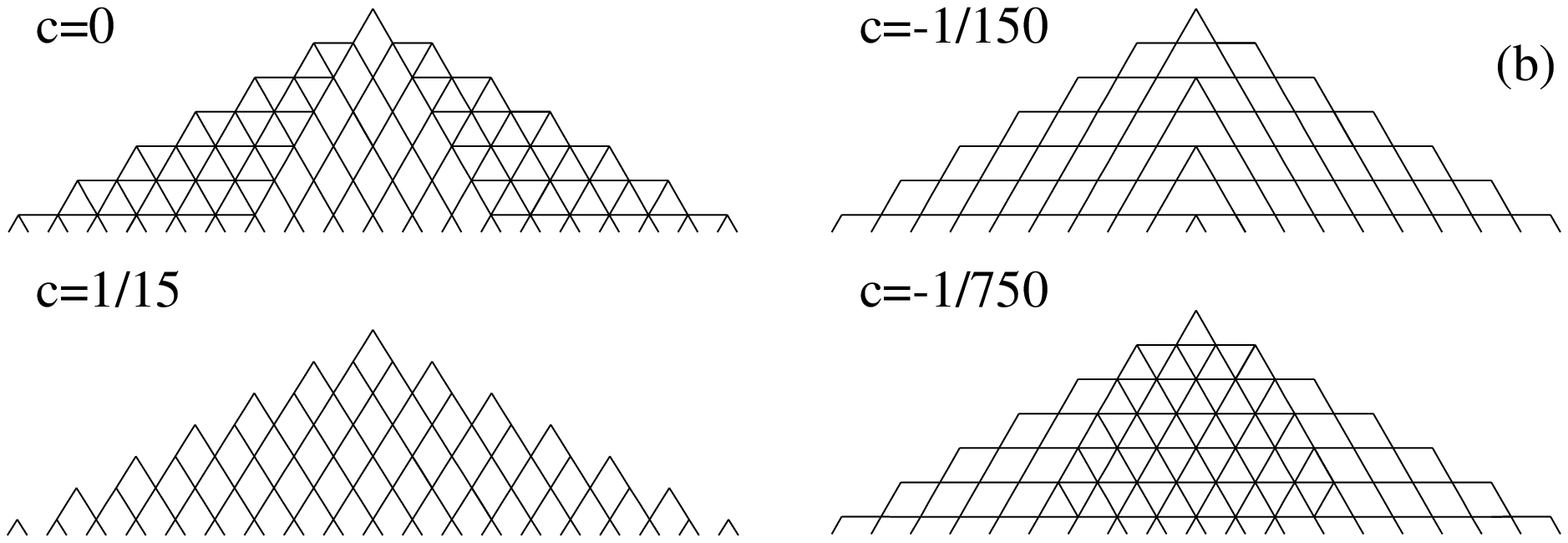,height=10.5cm,angle=-90,clip=}
\end{figure}
}
\def\PSFIGDB{
\begin{figure}[tbp]
\psfig{figure=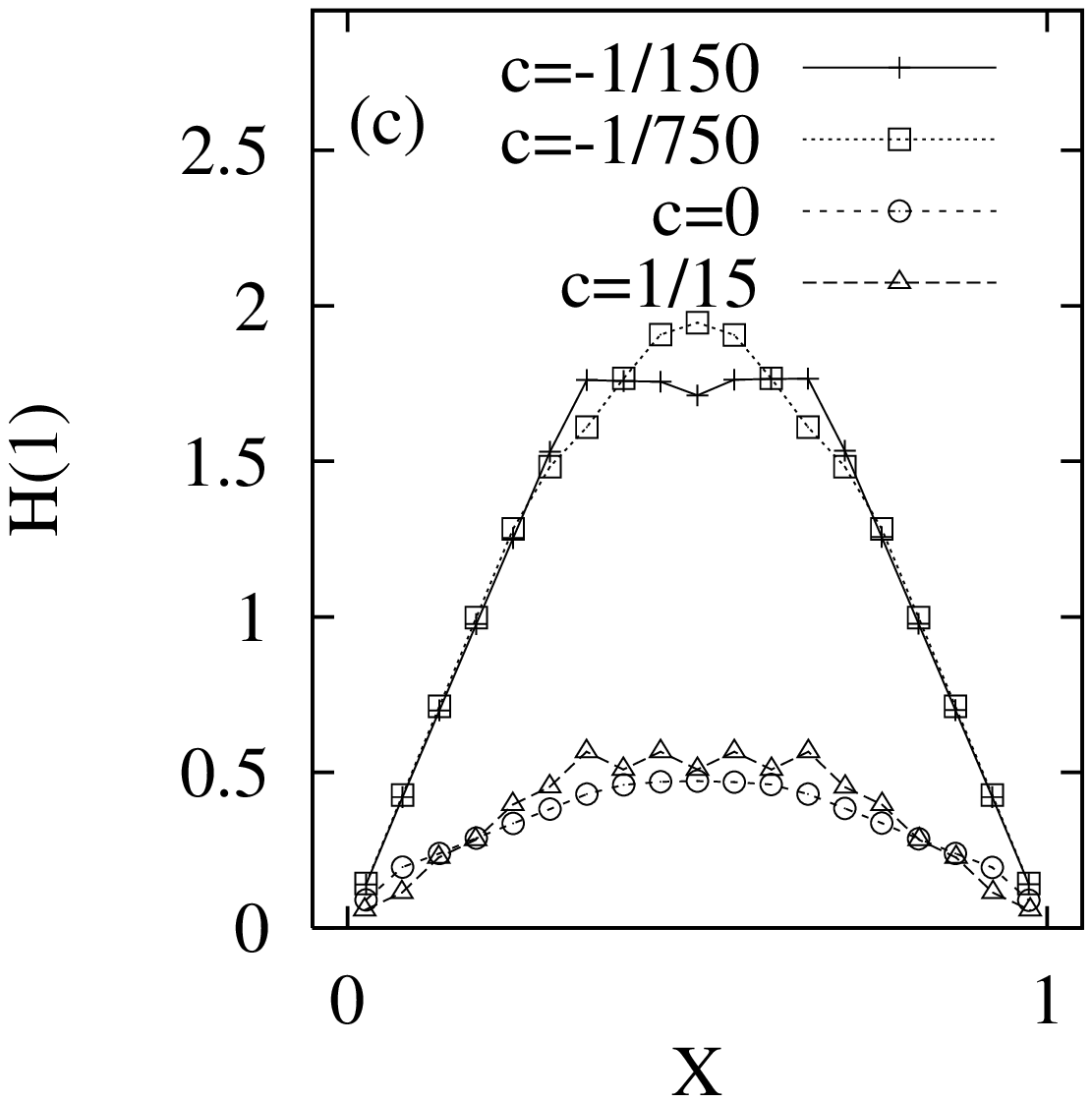,height=5.0cm,angle=-90,clip=}
\psfig{figure=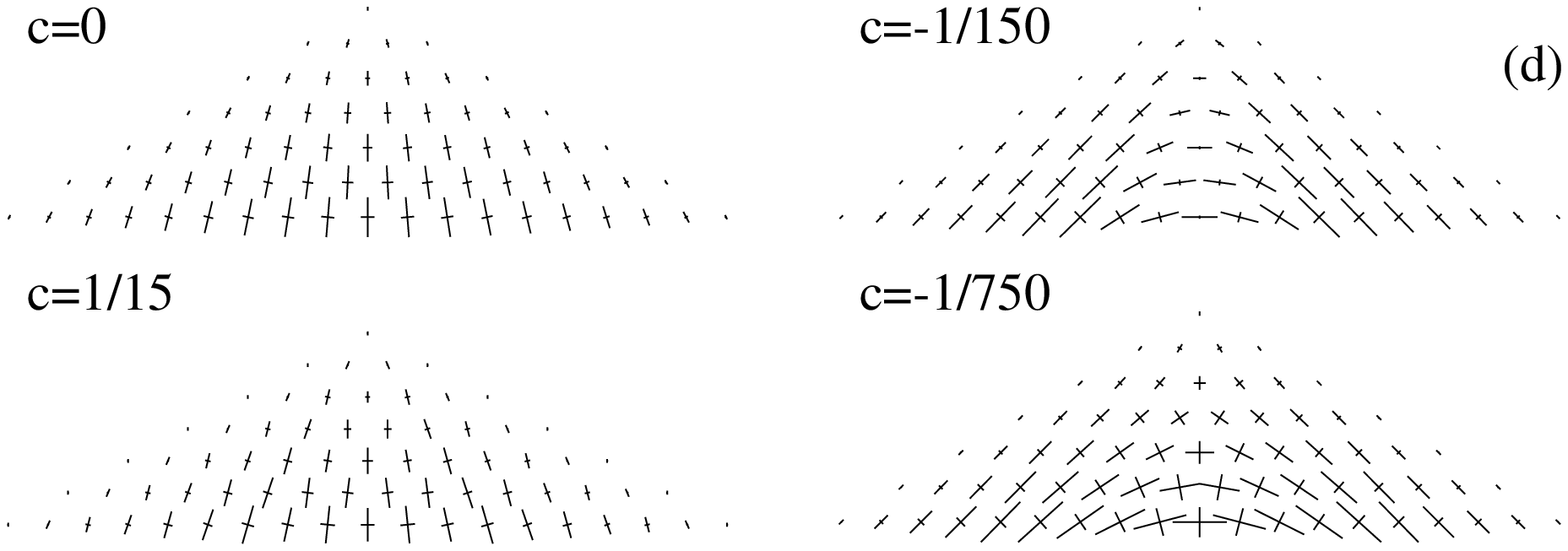,height=10.5cm,angle=-90,clip=}
\end{figure}
}

\def\PSFIGDC{
\begin{figure}[tbp]
\caption{
(a) Vertical stress $V(1)$, in row $M=1$, vs. $X$ for a $30^o$
pile with bumpy bottom and $L_1=19$. The immobile particles in row 
$M=0$ are separated by a distance $d_0 (1+c)$, i.e. are sqeezed 
together for negative $c$ or separated for positive $c$.
(b) The contact networks for the corresponding systems.
(c) Horizontal stress $H(1)$, vs. $X$.
(d) The principal axis of the stress tensor for the 
corresponding systems.
}
\label{fnet}
\end{figure}
}

\def\PSFIGE{
\begin{figure}[tbp]
\psfig{figure=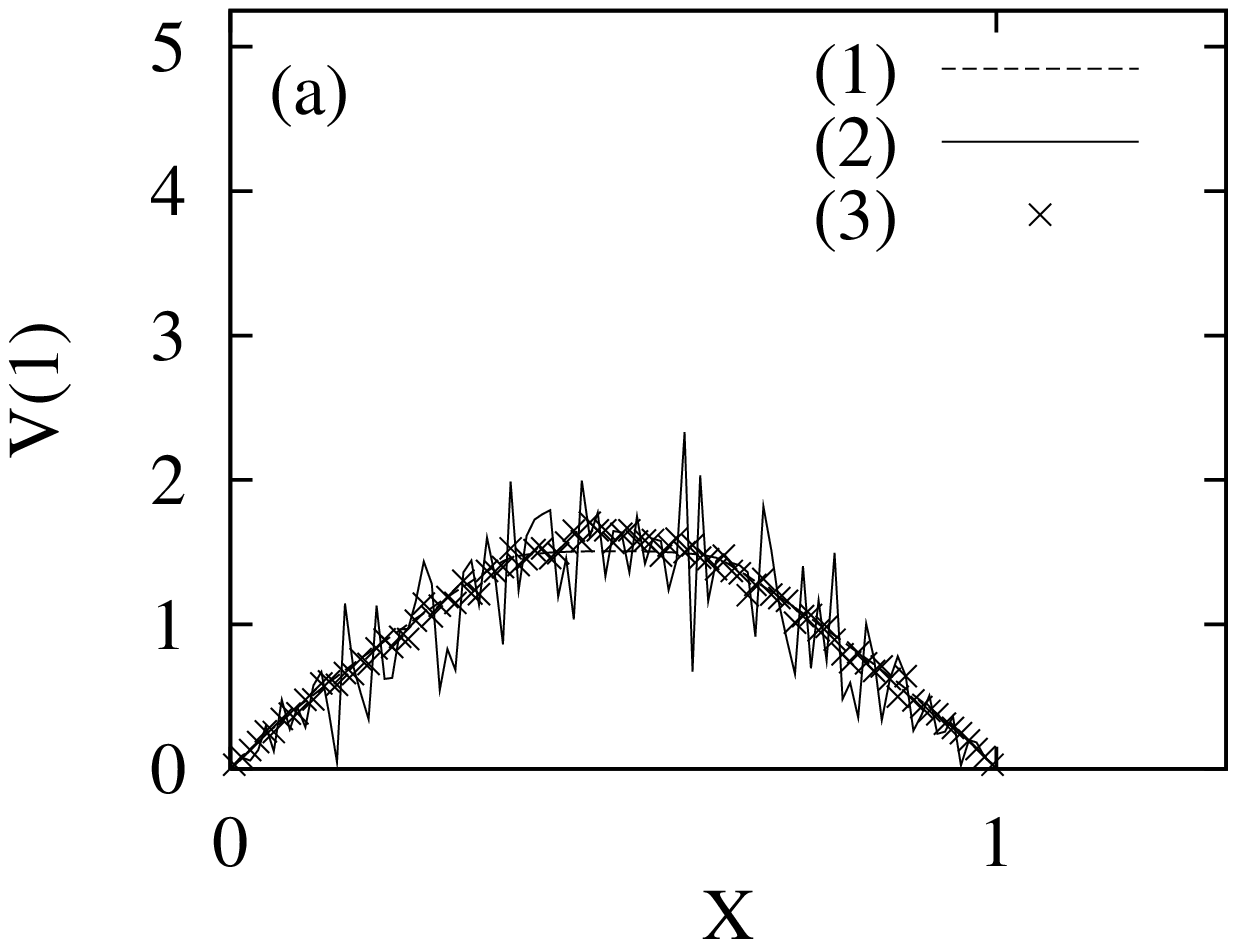,height=5.2cm,angle=-90,clip=}
\psfig{figure=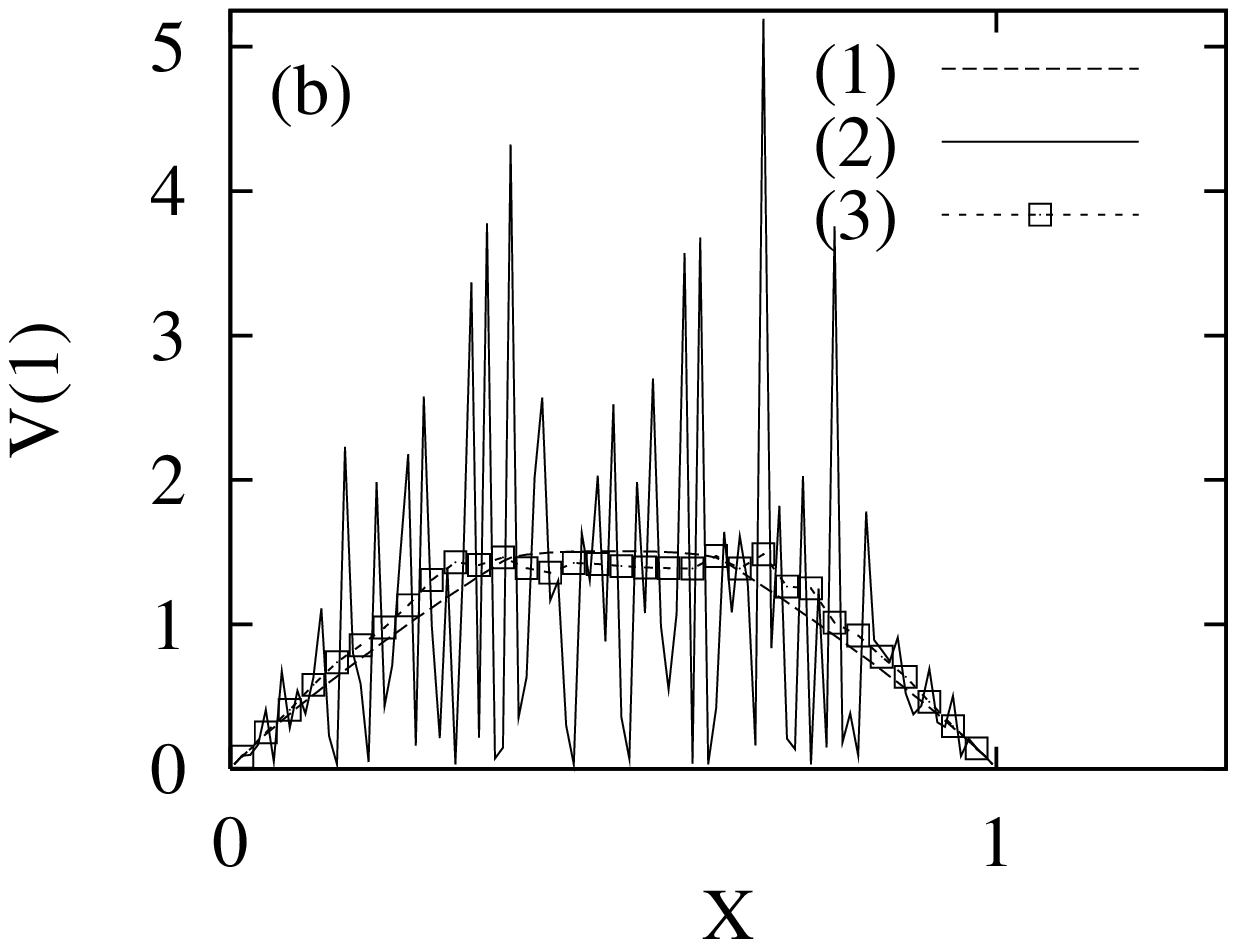,height=5.2cm,angle=-90,clip=}
\psfig{figure=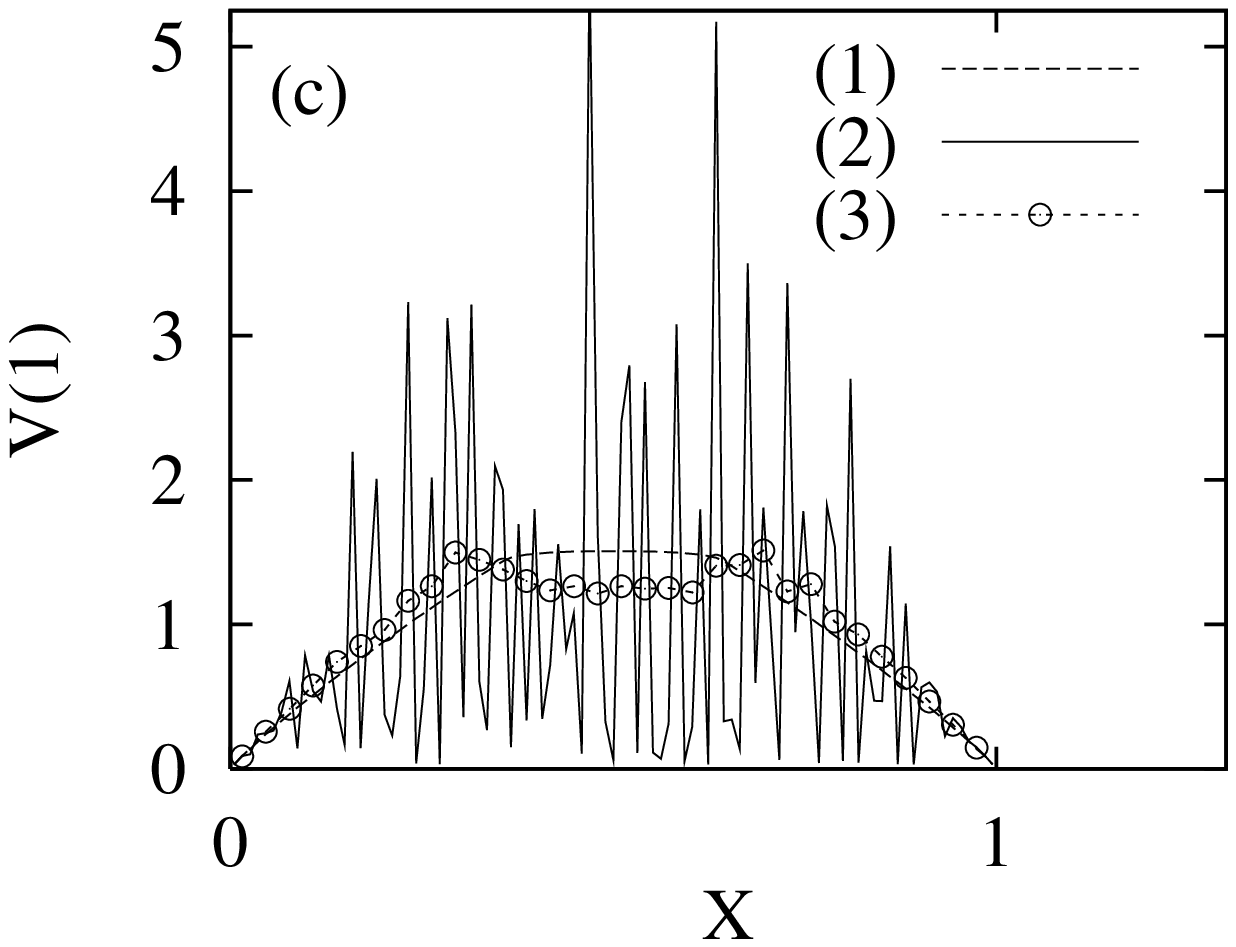,height=5.2cm,angle=-90,clip=}
\caption{
Vertical stress $V(1)$, in row $M=1$, vs. $X$ for a $30^o$
pile with bumpy bottom and $L_1=97$. The particle diameter is
homogeneously distributed in the interval $[d_0 (1-r/2), d_0 (1+r/2)]$.
The values of $r$ are $r=2/3000$ (a), $r=2/300$ (b), and
$r=1/30$ (c). The dashed line 
gives the result of Fig.\protect \ref{fig1}(b) with $r=0$ and $L_1=97$.
The solid line gives the result of one run and the symbols
correspond to an average over 40 runs for (a), or 100 runs and three
particles for (b) and (c).
}
\label{fdis}
\end{figure}
}

\def\PSFIGF{
\begin{figure}[tbp]
\psfig{figure=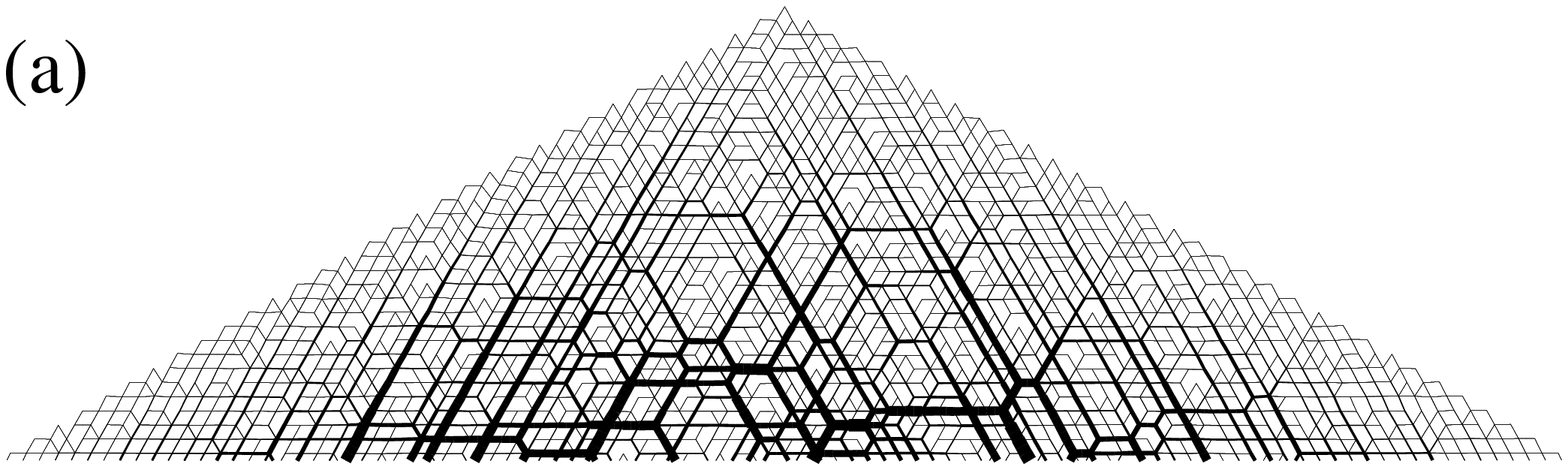,height=15cm,angle=-90,clip=}
\psfig{figure=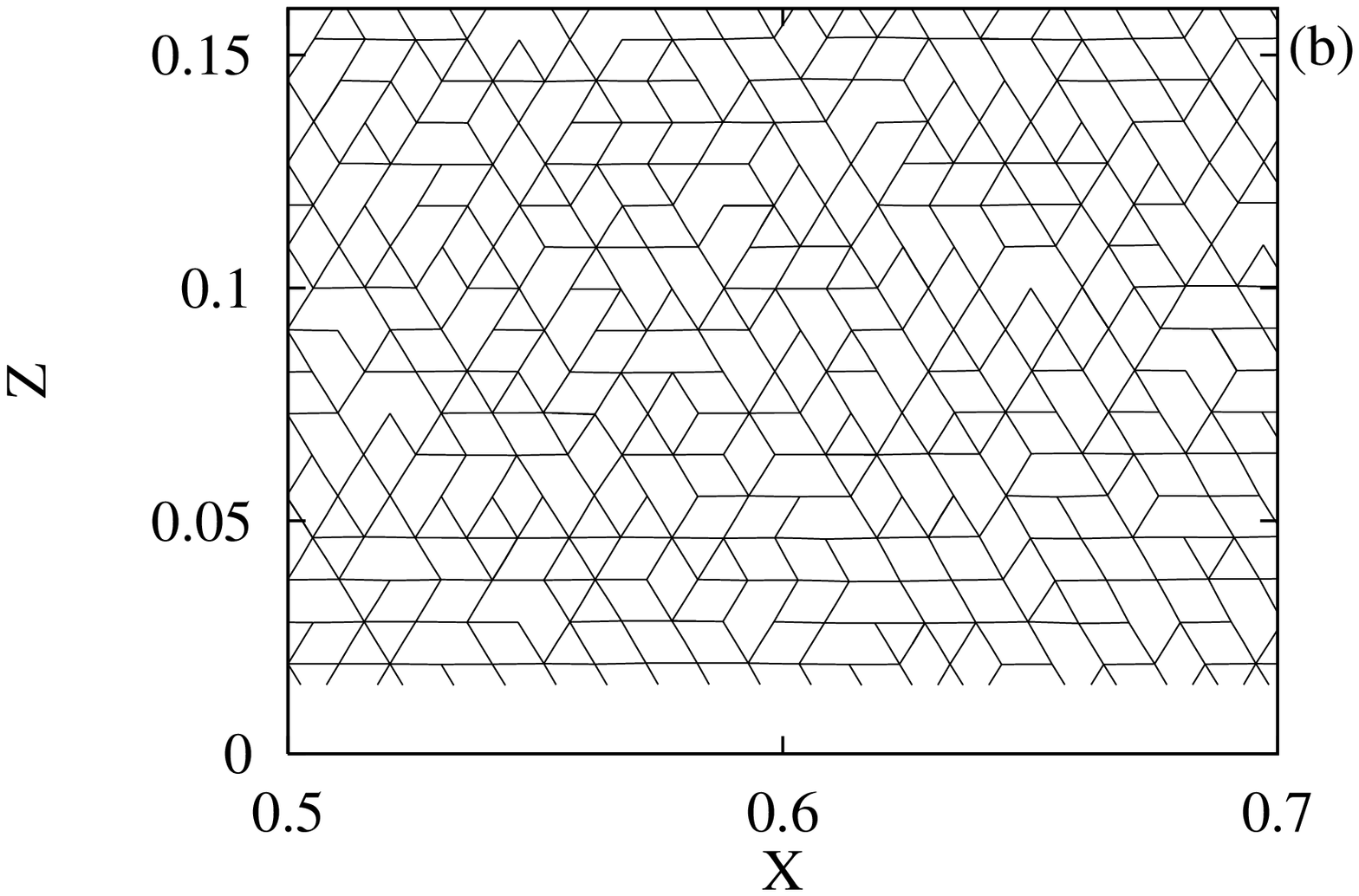,height=7.4cm,angle=-90,clip=}
\psfig{figure=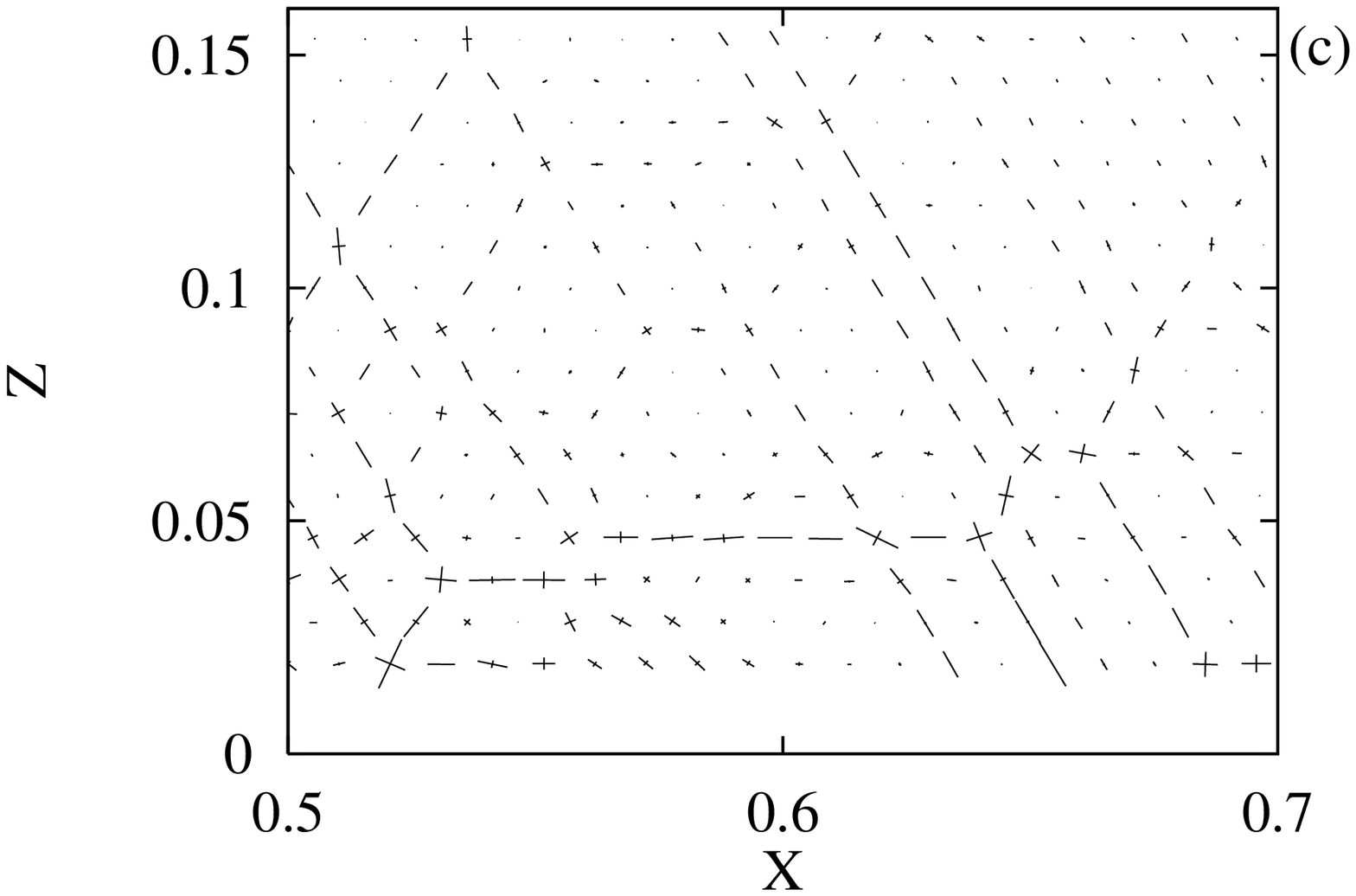,height=7.4cm,angle=-90,clip=}
\caption{
(a) Contact network of one pile from \protect{Fig.\ \ref{fdis}}(c).
    The line thickness indicates the magnitude of the contact force.
(b) Part of the contact network from (a).
(c) Principal axis of stress from the 
    same simulation as in (b).
}
\label{fdis2}
\end{figure}
}

\def\PSFIGG{
\begin{figure}[tbp]
\psfig{figure=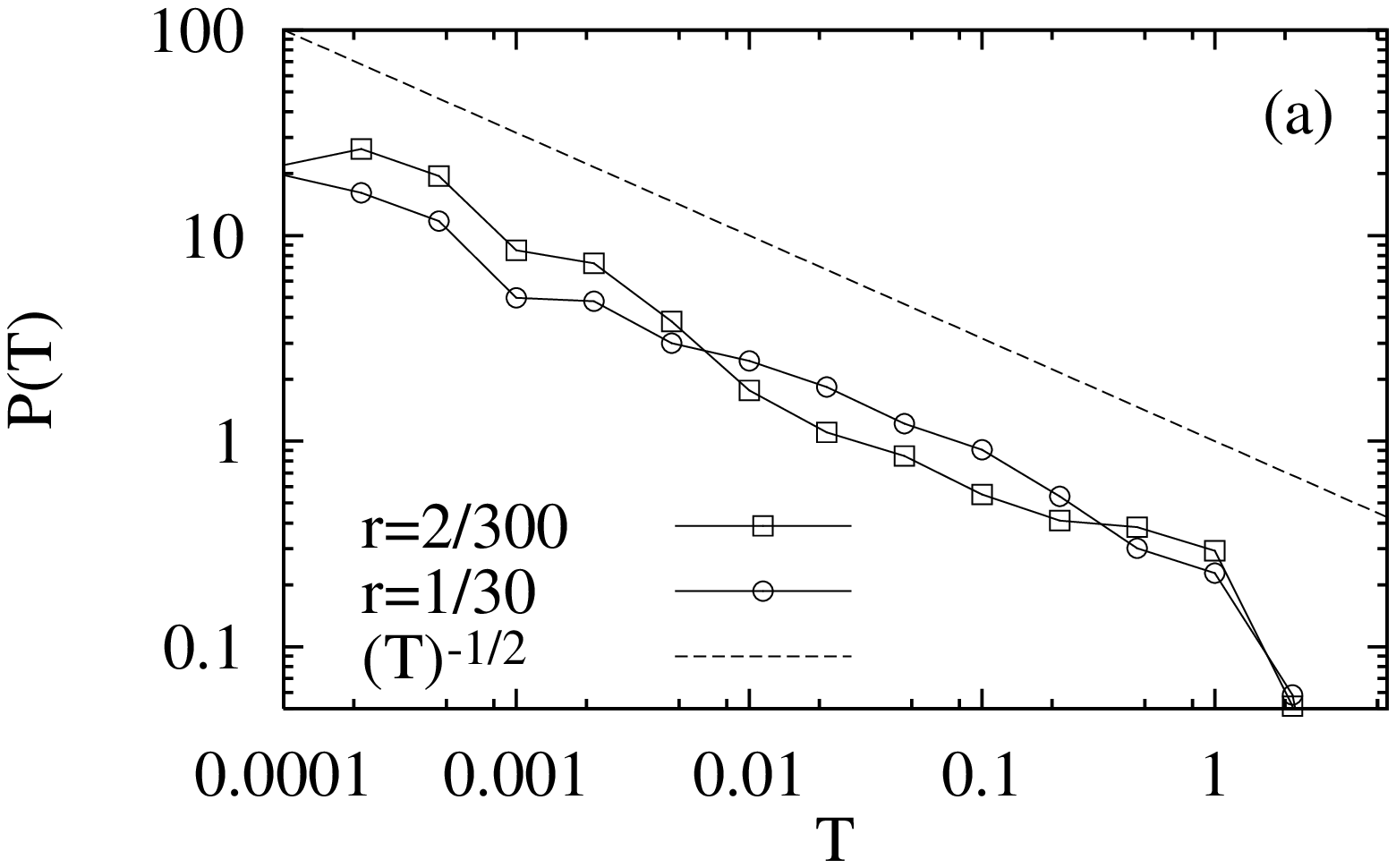,height=7.4cm,angle=-90,clip=}
\psfig{figure=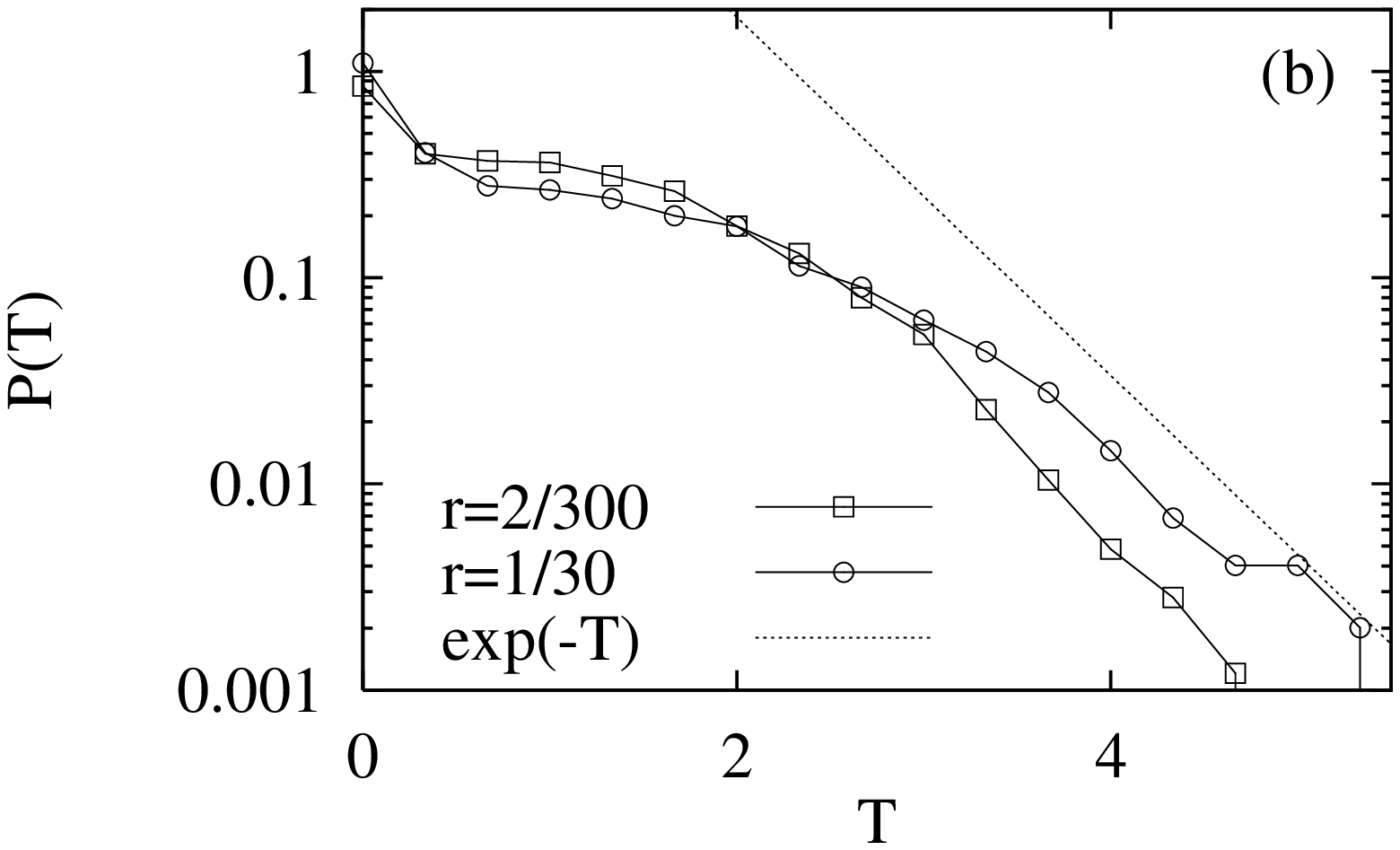,height=7.4cm,angle=-90,clip=}
\caption{
(a) Double logarithmic
    plot of the probability distribution of small vertical stresses 
    $T$ in row $M=1$. We skip 10 particles at the right and the left 
    and average over 100 runs. see Eq.\ \protect{\ref{eqtzz}}.
(b) Log-linear plot of the probability distribution for large stresses
    from the same data as in (a).
}
\label{fdis3}
\end{figure}
}

\def\PSFIGH{
\begin{figure}[tbp]
\psfig{figure=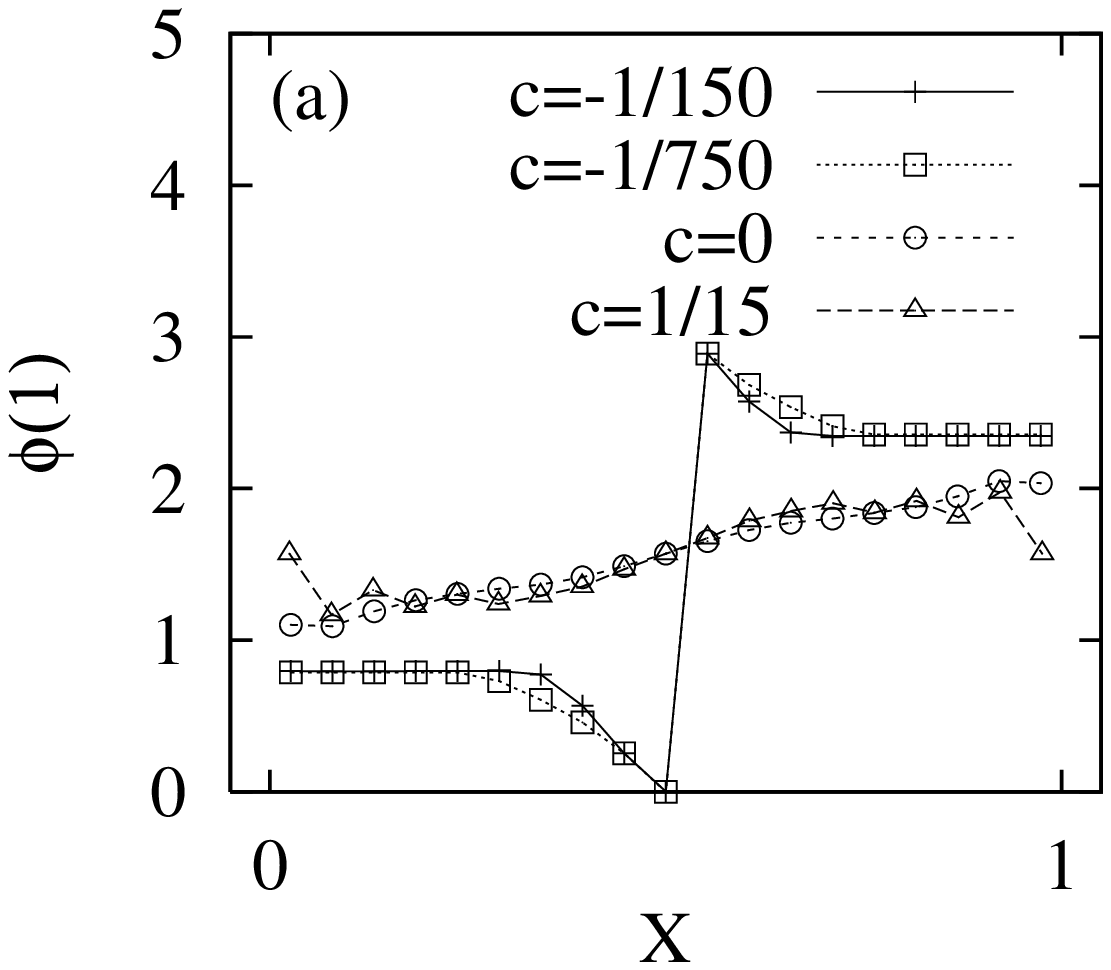,width=5.4cm,angle=-90,clip=}
\psfig{figure=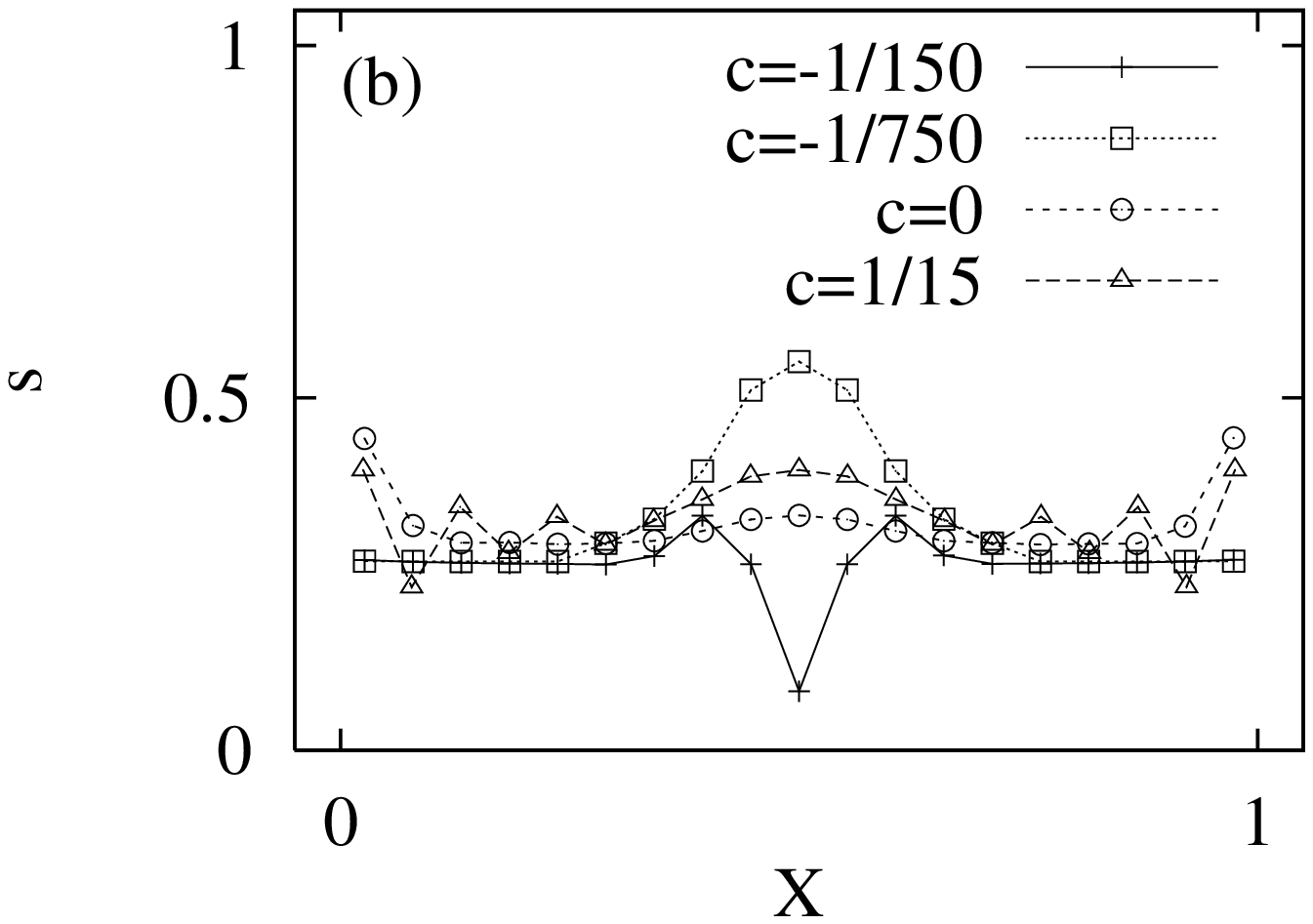,width=5.4cm,angle=-90,clip=}
\caption{
(a) The angle of the major principal axis of the stress, $\phi(1)$,
    in row $M=1$ vs. $X$, from the simulations in
    \protect{Fig.\ \ref{fnet}}.
(b) The ratio $s=S_{min}/S_{max}$ of minor to major principal axis vs. i
    $X$ from the simulations in (a).
}
\label{fphi}
\end{figure}
}

\section{Introduction}

In recent years the physics of granular materials has received
growing interest \cite{jaeger96}. One of the many interesting 
features of granulates is the stress distribution in static 
or quasi-static arrays. In contrast to a liquid, the pressure
in a silo, filled with e.g. grains,
is not increasing linearly with depth, but saturates
at a certain value \cite{janssen95}. This is due to internal
friction and due to arching, so that the walls of the silo carry
a part of the materials' weight. In sandpiles no walls are present so that
the situation may be different, i.e. the total weight of the pile
has to be carried by the bottom. However, the distribution of forces
under and also inside the pile is not yet completely understood. 
Experiments on rather large piles
show that the normal force has a relative minimum under the top
of the pile, the so-called dip \cite{trollope80,smid81}.
On a much smaller scale, the stress chains are observed,
i.e. stresses are mainly transported along selected paths
and the probability distribution of stress spans orders of magnitude
\cite{liu95,radjai96b,ouaguenouni97}.

One simple model pile is an array of rigid spheres, arranged 
on a diamond lattice, i.e. with four nearest neighbors each
\cite{liffman92,hong93}. The force under such a pile is
constant in contrast to the experimental observations, and also
periodic vacancies
in such a configuration do not lead to a dip in the pressure at the bottom
\cite{huntley93}. The variation of the size of some of the particles or
an attractive force between the particles may lead
to a non-constant force under the pile \cite{liffman94}. 
Continuum approaches 
\cite{edwards89,bouchaud95,edwards96b,wittmer96b,wittmer96}
may lead to a dip in the vertical stress if the correct
assumptions for the constitutive equations are chosen.
Edwards \cite{edwards89} introduced the notion that a 
pressure minimum can result from compressive stresses
aligning in fixed directions. Wittmer et al.
\cite{wittmer96b,wittmer96} embellished this
idea recently with concrete calculations in agreement with
the experimental data \cite{smid81}.
A lattice model based on a random opening of contacts 
\cite{hemmingsson96} also shows the dip in average over many 
realizations.

In this study we focus on 2D-situations, with particles on an
almost regular lattice, which we analyse using
MD-simulations. The aim is to find the dip under 
conditions as simple as possible and to understand the
stress networks and arches. We describe the simulation
method used in Sec.\ \ref{sec:md} and discuss the physics of 
particle contacts in Sec.\ \ref{sec:contacts}. The results are
presented in Sec.\ \ref{sec:results} and are discussed in 
Sec.\ \ref{sec:conclusion}.

\section{Simulation Aspects}
\label{sec:md}

The elementary units of granular materials
are solid ``mesoscopic'' grains, interacting on contact.
The surface is in general rough on a microscopic scale
and solid friction is usually found. Here, we focus on
properties of granular systems in the absence of friction.
We will examine in how far phenomena like stress chains and
arching depend on friction by neglecting solid friction.
However, we have some kind of ``geometrical friction'', since
the particles restrict the motion of their neighbors due
to excluded volume effects.

Without friction, energy may still be dissipated by
e.g. viscous deformations, modelled here by a 
simple viscous dashpot, active during the contact.

Since we are interested in static arrangements of
particles in the gravitational field, we use strong viscous
damping, in order to reach the steady state quickly.
For the relaxation of the array we use a molecular
dynamics (MD) procedure 
\cite{cundall79,allen87}, in order to allow contacts to break.
The MD method is not the best choice for a fast relaxation,
but closing and opening of contacts is implemented straightforwardly.

\subsection{Initial and Boundary Conditions}

In the simulations $N$ spherical particles, with diameters
$d_i$, ($i$ = 1,...,$N$) are used. If not explicitly mentioned
we use monodisperse spheres of diameter $d_i = d_0 = 1.5$mm.
The $N$ particles are placed into a container
with different boundary conditions at the bottom and also different
system sizes. Starting from a regular
close-packed triangular arrangement with $L$ particles
in the lowermost layer $M=0$ at the bottom, we model heaps of slope
$60^o$ or $30^o$ by adding $L_M=L-M$ or $L_M=L-3M$ particles for layer $M$
respectively. The number of particles is thus $N^{(60)}=H^{(60)}(L+1)/2$ or
$N^{(30)}=H^{(30)}(L-3(H^{(30)}-1)/2)$ with the number of layers $H^{(60)}=L$ 
or $H^{(30)}=$int$[(L-1)/3]+1$. The largest pile we simulate has 
$L=100$ and thus $N^{(30)}=1717$.

The initial velocities and overlaps of the particles are set to zero 
if not explicitly given, gravity is slowly tuned from zero to 
the selected magnitude and the system is simulated until the
kinetic energy is several orders of magnutide smaller than the 
potential energy, and the stresses no longer vary. 
The particles at the bottom
layer $M=0$ are either fixed, or may slide horizontally and penetrate
the bottom vertically. In the sliding case, only the outermost
particles are fixed in horizontal direction by the sidewalls.
For a schematic drawing of the four possible situations see
Fig.\ \ref{fig0}(a). The possible configurations of a regular
contact network are schematically drawn in Fig.\ \ref{fig0}(b).
\PSFIGO

\subsection{Molecular Dynamics Method}

For the integration of the equations of motion
we use a fifth order predictor-corrector MD-scheme, 
see Ref.\ \cite{cundall79,allen87}.
Since we are interested in a static situation in 2D, 
with almost monodisperse particles, no particle
has more than six nearest neighbors. For the simulation we keep the neighbors
in memory in order to reduce the computational effort.

There are two forces acting on
particle $i$ when it overlaps with particle $j$, i.e. when the distance 
$r_{ij} = \mid \vec r_j - \vec r_i \mid \le (d_i + d_j)/2$.
We use an elastic force
\begin{equation}
\vec f^{(i)}_{el} = k ( r_{ij} - {1 \over 2}(d_i
+ d_j)) \vec n_{ij} ,
\label{fel}
\end{equation}
with the spring constant $k$, acting on particle $i$ in normal direction
$\vec n_{ij} = (\vec r_{j} - \vec r_{i}) / r_{ij}$.
The second force in normal direction is dissipative  
\begin{equation}
\vec f^{(i)}_{diss} = \mu (\vec v_{ij} \cdot \vec n_{ij})
\vec n_{ij} ,
\label{fdiss}
\end{equation}
accounting for the inelasticity of the contacts. 
In Eq.\ \ref{fdiss} the constant 
$\mu$ is a phenomenological dissipation coefficient, and
$\vec v_{ij} = \vec v_j - \vec v_i$ is the relative velocity of
the particles $i$ and $j$.
As mentioned above, we neglect tangential forces.
The contact of a particle with a wall or an immobile particle
is mimicked by setting the mass of the immobile contact partner
to infinity. Finally, the influence of gravity
$g$ is readily included into the equations of motion.

\section{Contacts}
\label{sec:contacts}

Now we are interested in the static limit, where particles ideally
have zero relative velocities and are either in contact or separated
by a gap and in the latter case do not interact. However, we will 
discuss in this section the dynamics of contacts in order to estimate
the typical scales of the system.

\subsection{Two particle contacts}

Since we use no tangential forces, we will discuss the normal
direction of a contact only. Considering the collision of two particles,
the situation is modelled 
by a spring and a dashpot, see Eqs.\ \ref{fel} and \ref{fdiss}
so that the relative acceleration during contact is
$y''~= (d^2/dt^2) r_{ij} = f^{(j)}/m_j-f^{(i)}/m_i$, with
$f^{(i)} = f^{(i)}_{el} + f^{(i)}_{diss}$.
Due to force balance we set $f^{(j)} = - f^{(i)}$ what
leads to a differential equation for negative
penetration depth $y=r_{ij} - (1/2) (d_i + d_j)$:
\begin{equation}
y''~+~2 \gamma y'~+~\omega_{0}^{2} y~=~0 .
\label{eq:omega0}
\end{equation}
In Eq.\ \ref{eq:omega0}, $\omega_0=\sqrt{k / m_{ij}}$,
$\gamma = \mu /(2 m_{ij})$, and the
reduced mass $m_{ij}=m_i m_j / (m_i + m_j)$. The solution of 
Eq.\ \ref{eq:omega0} is for $y \le 0$:
\begin{equation}
y(t)~=~\left ( {{v_{0}} / {\omega}} \right )~\exp(-\gamma t)~\sin(\omega t) ,
\label{eq:x0}
\end{equation}
with the corresponding velocity:
\begin{equation}
y'(t)~=~( {{v_{0}} / {\omega}} ) ~\exp ({- \gamma t})~
 \left [ { - \gamma ~\sin(\omega t) + \omega ~\cos(\omega t)~} \right ] .
\label{eq:x1}
\end{equation}
In Eqs.\ \ref{eq:x0} and \ref{eq:x1} 
$v_0 = y'(0)$ is the relative velocity before collision 
and $\omega=\sqrt{\omega_{0}^{2}-\gamma^{2}}$ the damped frequency.
As long as $\gamma^2 < \omega_0^2$, the typical duration of 
the contact of two particles is:
\begin{equation}
t_c = {{\pi} / {\omega}} ,
\label{eq:tc}
\end{equation}
because the interaction ends when $y(t) > 0$.
The coefficient of restitution $\epsilon$ is defined as
the ratio of velocities after and before contact
$\epsilon = - y'(t_c)/y'(0)$ so that Eqs.\ \ref{eq:x1} and \ref{eq:tc} 
lead to
\begin{equation}
\epsilon =  \exp ({-{ \pi \gamma / \omega } }).
\end{equation}
From Eqs.\ \ref{eq:x0} and \ref{eq:x1} the maximal penetration 
depth $y_{max}$ follows
the condition $y'(t_{max})=0$, so that 
$\omega t_{max} = \arctan(\omega / \gamma) = \arcsin(\omega / \omega_0)$
and
\begin{equation}
y_{max} = ({ v_0 / \omega }) \exp({-{ \gamma t_{max} } }) 
           \sin( \omega t_{max} )
        = ({ v_0 / \omega_0 }) \exp \left [ ({-{ \gamma / \omega })
           \arcsin( \omega / \omega_0 ) } \right ] .
\label{eq:xmax}
\end{equation}
The maximum penetration depth $y_{max}(v_0)$
is in the case of, say, steel particles
much smaller than the particle diameter. Thus we check in our simulations
that $y_{max}$ is always orders of magnitude smaller than $d_0$.

The elasticity $k$ in Eq.\ \ref{fel}
is e.g. a function of the Young modulus and the Poisson ratio,
which are material dependent and thus fix $t_c$ for a given material
in our simplified model. Using the theory of Hertz, a more
complicated dependence of $k$ on the impact velocity, the 
elasticity , and the penetration depth is found, 
e.g. $k \propto y^{1/2}$. In Ref. \cite{luding94d}, the contact
time of two steel spheres with diameter $d = 1.5$mm and with 
an impact velocity of $v_0 = 1$m/s
was evaluated to $t_c \approx 4.6 \times 10^{-6}~$s. 
We checked for some situations that the more realistic Hertz model 
does not change the results \cite{luding97b} and thus used the
simpler linear model. For a detailed discussion of different
MD models and force-laws see Ref.\ \cite{schafer96}.

For weak dissipation
$t_c$ is proportional to $k^{-1/2}$, so that an increase
of $k$ by a factor of 100 decreases $t_c$ by a factor of 10. Now taking
physical values for $t_c$ leads to extremely long MD-
computing times for a given simulation time. One has to insure 
that the time scales of the system, i.e. $t_c$, and of the
algorithm, i.e. the integration time step $t_{MD}$, are well separated. 
Ideally one should have $t_{MD} << t_c$. The MD-simulations
reported here were done with $t_{MD} < t_c / 40$.
Using contact times in the range $10^{-3}$s$ < t_c < 10^{-5}$s, by
choosing $k$ according to Eq.\ \ref{eq:tc}, we have 
simulation time steps in the range $2.5 \times 10^{-3}$s
$ < t_{MD} < 2.5 \times 10^{-7}$s.

In our simulations we have as a typical set of parameters 
$d_0 = 1.5$mm, $k/m_{ij}= 6.67 \times 10^6$s$^{-2}$,
$\gamma = 1.67 \times 10^3$s$^{-1}$ and $t_{MD} = 10^{-5}$s. 
These parameters lead with the above
equations to $t_c = 0.97 \times 10^{-3}$ s, and
$\epsilon = 0.2$, i.e. rather strong dissipation.

\subsection{Multi particle contacts}

In Refs.\ \cite{luding94d,luding94b,luding95} the above defined
interaction law has been tested in the case of many particles in
contact at the same time. For the viscous interaction law, i.e.
the linear spring-dashpot model, energy dissipation is very
inefficient, i.e. the so-called ``detachment effect'' occurs.
The time a wave needs to propagate through a system
of size $l = L d_0$ was found to be comparable to $L t_c$.
Thus we will measure time in units of $L t_c$, and velocities 
in units of $d_0 / t_c$, i.e. particle size divided by the contact time, 
rather corresponding to the speed of sound inside the elastic material.
Thus we have the length $l$ as the product of 
typical time and typical velocity.

Dividing Eq.\ \ref{fel} by $kl$ we find that the dimensionless
deformation $x/l$ in a static situation, i.e. $y' = 0$, is proportional 
to the dimensionless elastic force $f_{el}/(k l)$. In the 
gravitational field the elastic force $f_{el}$ scales with $m g$, 
where $m$ is the mass of the pile, so that $m g \propto k l$. We
tested for several situations that our results do depend rather on 
the ratio $g/k$, than on the specific values chosen for $g$ or $k$.

\subsection{Stress Tensor and Scaling}

An important quantity that allows insight into the state of
the system is the stress tensor $\sigma$ \cite{goddard86,bathurst88}, 
which we identify in the static case with
\begin{equation}
\sigma^{(i)}_{\alpha \beta} = (1/V^{(i)}) \Sigma ~  q_\alpha f_\beta,
\end{equation}
where the indices $\alpha$ and $\beta$ indicate the 
coordinates, i.e. $x$ and $z$ in 2D, see Fig.\ \ref{fig0}. 
This stress tensor
is an average over all contacts of the particles within 
volume $V^{(i)}$, with $q$ denoting the distance between the 
center of the particle and the contact point, and $f$ denoting
the force acting at the contact point. Throughout this study we 
average over the contacts of one particles $(i)$ to get the stresses
for one realization.

From a static configuration of ``soft'' particles we may now calculate
the components of the stress tensor $\sigma_{xx}, \sigma_{zz},
\sigma_{xz}$, and $\sigma_{zx}$ and also define $\sigma^+ = (
\sigma_{xx} + \sigma_{zz})/2$, 
$\sigma^- = (\sigma_{xx} - \sigma_{zz})/2$,
and $\sigma^* = \sigma_{xz}$. 
Since we neglected tangential forces the particles are
torque-free and we observe only symmetric stress tensors,
i.e. $\sigma_{zx} = \sigma_{xz}$. The eigenvalues 
of $\sigma$ are thus $\sigma_{max,min} = \sigma^+ \pm 
\sqrt{(\sigma^-)^2 + (\sigma^*)^2}$, and the major eigenvalue
is tilted by an angle
\begin{equation}
\phi = \arctan \left ({\sigma_{max}-\sigma_{xx} \over \sigma_{xz}} \right )
     = {\pi \over 2} + {1 \over 2} 
       \arctan \left ({ 2 \sigma_{xz} \over {\sigma_{xx}-\sigma_{zz}} }\right )
\end{equation}
from the horizontal in counter clockwise direction.

In order to find the correct scaling for the stress we assume
like Liffman et al. \cite{liffman92,liffman94},
as a simplified example, a rigid triangle with the density $\rho$, the 
width $l$, the height $h$, and the mass $m = \rho h l / 2$.
Since the material is rigid, we find a constant force at the 
supporting surface, so that the pressure is also constant $p = 
m g / l = \rho g h / 2$. Thus we will scale the stress by the 
pressure $p$ and furtheron use the dimensionless stress 
\begin{equation}
S = { 2 \sigma \over \rho g h }
  = { \sigma l \over m g } 
  = { \sigma 2 a \over h m g },
\label{defs}
\end{equation}
with the volume $a = h l / 2$ of the triangular pile.
The vertical component will be abbreviated with $V=S_{zz}$,
the horizontal component with $H=S_{xx}$, and the shear
component $Q=S_{xz}$.
Besides the components of $S$ we will also plot the stress tensor in its
principal axis representation, i.e. for each particle
we plot the scaled major principal axis along $\phi$ and the minor
axis in the perpendicular direction.

\section{Results}
\label{sec:results}

\subsection{Piles with Bumpy Bottom}

\subsubsection{Comparison of Piles with different Slope}

The first situation we address is a homogeneous pile, as 
assumed in Refs.\ \cite{liffman94,hong93}. 
Here we use $L_1=20$ particles
in row $M=1$ and create a $60^o$ pile. The $L_0=21$ particles
in the lowermost row $M=0$ are fixed with separation $d_0$.
The particles have no horizontal contacts, so that the
contact network is a diamond structure. As predicted
in Refs.\ \cite{liffman94,hong93} the normal force at
the bottom is a constant, independent from the horizontal
coordinate. In Fig.\ \ref{fig1}(a) we plot the components of
the dimensionless stress tensor $S(1)$ versus $X=x/l$ for 
the lowermost row of mobile particles, $M=1$. The vertical component
is constant, and due to the scaling used $V = 1$.
We compare this result with two $30^o$ piles with either
$L_1=20$ or $L_1=97$ and plot again $S(1)$ vs. $X$ for both 
system sizes in Fig.\ \ref{fig1}(b). 
For the $60^o$ pile the diagonal elements
of $S$ are constant, whereas for the $30^o$ piles we observe
a plateau in the center with decreasing stresses towards the
left and right ends of the pile. Our simulation results are in 
agreement with analogous simulations in Ref.\ \cite{liffman94}, i.e. 
we observe no sharp edges in the stresses, where the slopes change,
 as predicted by the theory in Ref.\ \cite{liffman94}. 

\PSFIGA

From Fig.\ \ref{fig1} we conclude that our soft particle
model is able to reproduce the known analytical results of
Refs.\ \cite{liffman94,hong93}. $V=1$ corresponds to 
the constant normal stress $\sigma_{zz} = mg/l$ and thus to the 
normal force $f_z$ exerted on each particle in row $M=1$. 
Here $f_z = d_0 \sigma_{zz} = mg/L$, with the
mass of the pile $m = L(L+1) m_0/2$, and the mass of one particle 
$m_0$. Our result $f_z = (L+1) m_0 g / 2$ coincides with 
Ref.\ \cite{liffman94} [see eq.42 therein]. 

\subsubsection{Variation of System Width for Bumpy Bottom}
\label{sec:varb}

In this subsection we will examine the difference between
the theoretical predictions for the stresses and the numerical 
simulations, both in Ref.\ \cite{liffman94} for the $30^o$ pile. 
The theory is based on
the assumption that the contact network is a diamond lattice.
Thus we perform different simulations with a $30^o$ pile with 
$L_1=19$ and change the contact network by increasing or decreasing
the separation of the fixed particles in row $M=0$. The centers of
the particles in the lowermost row are separated by a distance 
$d_0 (1+c)$, with the $c$ values $c$ = 1/15, 0, -1/750, and -1/150.
In Fig.\ \ref{fnet}(a) and (c) we plot the vertical and horizontal
components of the stress tensor, and in Fig.\ \ref{fnet}(b) and (d)
we plot the contact network and the principal axis of the stress
tensor respectively. 

\PSFIGDA
\PSFIGDB
\PSFIGDC

The interesting result is that the vertical stress 
in Fig.\ \ref{fnet}(a) has a dip for negative $c$ values, 
the depth of which increases with increasing magnitude 
of $c$ \cite{liffman94}. The horizontal stress in 
Fig.\ \ref{fnet}(c) is much larger for negative $c$ as
for positive $c$.

From Fig.\ \ref{fnet}(b) we observe that the assumption of 
a perfect diamond lattice
for the contacts is true only for $c = 1/15$. The vertical
stress $V(1)$ has a zig-zag structure that we relate 
to the steps at the surface of a $30^o$ pile. For the naively
used $c = 0$ and also for small negative $c = -1/750$ we have
a contact network with regions of coordination number 4 and 6, 
corresponding to the triangular or the diamond contact network.
For sqeezed bottom particles, i.e. $c = -1/150$, the contact
network is again a diamond lattice, but the orientation is 
tilted outwards from the center.
From Fig.\ \ref{fnet}(d) we evidence arching for negative
$c$ and no arching for positive $c$. Seemingly, a tilted 
diamond lattice is neccessary for an arch to form in 
this situation.

In Fig.\ \ref{fphi}(a) we present for the simulations
from Fig.\ \ref{fnet} the angle $\phi(1)$, see Eq.\ \ref{fphi}
about which the major principal 
axis is rotated from the horizontal in counterclockwise direction.
For $c < 0$ we observe a constant angle in the outer part [consistent
with the fixed principal axis (FPA) theory in Ref.\ \cite{wittmer96}],
and a transition region in the center. We observe FPA only for
negative $c$ when we also find arching.

In contrast, for $c \ge 0$ we observe a slow continuous variation
of $\phi(1)$ over the whole pile. In Fig.\ \ref{fphi}(b) we plot 
the ratio of the principal axis $s=S_{min}/S_{max}$ and observe an
almost constant value in the outer region of the pile,
whereas in the inner part the ratio is strongly $c$ dependent.
\PSFIGH
From a detailed comparison of the contact network and the stress
tensor we may correlate several facts: Firstly, the ratio of
the principal axis, $s$, seems to determine whether
the contact network is a triangular or a diamond structure, 
the latter with one open contact. 
For $c=0$ and for $c=-1/750$ we observe the triangular 
contact network if $s$ is large.
Secondly, the direction of the diamonds is correlated to $\phi$, 
i.e. we observe the tilted diamond lattice (for negative $c$) if 
the major axis is tilted far enough from the horizontal.

\subsubsection{Removing Particles from the Pile}

Now we use the pile from Fig.\ \ref{fig1}(a), i.e. $60^o$ with $L_1=20$,
and examine the influence of one removed particle on the stress
distribution. Here we remove the third, fifth, and seventh particle
denoted with $R$ = 3, 5, and 7 respectively, from the right in row $M=7$.
We relax the pile and plot the
vertical normalized stress in row $M=1$, i.e. at the bottom,
in Fig.\ \ref{figt}(a). Evidently, the formerly constant stress of the
complete pile (solid line) is disturbed. We observe that the
stress decreases in the region below the missing particle at 
$X$ = 0.55, 0.65, and 0.75
for $R$ = 7, 5, and 3 respectively. Interestingly, the stress
is minimal when following lines parallel to the slopes of the
pile, towards row $M$=1, starting from the vacancy.
Note that following the slopes means here: following a line
in the diamond contact network. The lines of contact are here 
tilted by $60^o$ from the horizontal and thus are parallel to the
slopes. Going from the minimum value outwards
we observe a sudden jump to the maximum value of $V(1)$.
When a particle close to the center of the pile is removed, i.e. $R$
= 7, for $L_7 = 14$, 
the stress pattern is almost symmetric to the center $X=0.5$, whereas the
pattern gets more and more asymmetric with decreasing $R$.

When a particle is removed, this particle can not longer transfer
the stresses to its lower neighbors. Therefore the minimum stress
is found when following the slopes starting from the missing
particle. The stress which has not been carried by the missing
particle has thus to be transferred along its right and left
neighbors, what leads to the maximum stresses just outwards from
the minimum stresses.
\PSFIGT
In order to clarify this result we plot the vertical stresses
inside the pile at different heights $M$ = 1, 4, 7, 10, and 13
in Fig.\ \ref{figt}(b). With increasing $M$, i.e. increasing height in the
pile, $S$ decreases since the weight of the part of the pile above $M$
decreases. 
Inside the pile, the stress is
minimal when following the slopes downward, starting from the missing
particle. Interestingly, we observe an asymmetric stress also for 
$M > 7$. In Fig.\ \ref{figt}(c) we plot the contact network
for the case where particle $R=7$ is removed from row $M=7$.
We observe an increase of the number of contacts only for the
neighbors of the vacancy. From the principal axis of stress,
in Fig.\ \ref{figt}(d), we observe an arch-like structure, i.e.
the stress below the vacancy is comparatively small. Furthermore,
the direction of the major principal axis is almost vertical
below the vacancy, and tilted outwards for the particles which 
carry larger stresses.

We learn by removing one particle from the pile, that stress 
decreases below the vacancy; however, the minimum of stress
is observed when following the internal structure of the pile
downwards, i.e. lines tilted by $60^o$ from the horizontal. Much larger
stresses are observed outwards from the minima in stress, i.e. 
an arch-like structure is found already for one missing particle.

Note that this simulation is not in contradiction to the 
discussion, concerning point source terms, in Ref.\cite{wittmer96}.
Wittmer et al. discuss the change an infinitesimally small mass
element has onto the stress distribution and conclude that
the (small) weight is propagated along ``rays'' mainly into the direction
of gravity. In our case the mass removed is quite large and thus 
the contact network is deformed what leads to the different effects 
described above.

\subsection{Pile with Smooth and Flat Bottom}

In contrast to the piles with bumpy bottom, corresponding to the
limit of very large friction, we model now a pile on a smooth and
flat bottom, i.e. the limit of no friction. Note that this situation 
is stable only if the outermost particles are fixed.

\subsubsection{Comparison of Piles with different Slopes}

The next situation we describe is a pile on a flat, smooth
bottom, i.e. the particles in row $M=0$ are allowed to move.
Only the left- and rightmost particles are fixed horizontally 
by the corresponding wall. In Fig.\ \ref{fig2}(a) we show 
the results for two $60^o$ piles with $L_0=20$ and $L_0=40$.
The vertical component $V$ of the stress is not constant
and the horizontal component $H$ is getting very large
close to the walls, since vertical stresses are transferred into
the horizontal direction and propagate directly outwards in row $M=0$. 
In the case of $L_0=40$ we observe a relative minimum of the 
vertical stress in the center, $X=0.5$. 
\PSFIGB
In Fig.\ \ref{fig2}(b) we compare the result of Fig.\ \ref{fig1}(b), 
i.e. $L_1=97$ to situations on smooth and flat bottom with
$L_0=22$ and $L_0=100$. The dashed lines give the vertical stress
in row $M=1$ (c) and $M=0$ (d). We observe fluctuations
at the shoulders of the pile and again a dip in the 
center of the pile, $X=0.5$. In order to find an explanation 
for this behavior we plot the contact networks in Figs.\ \ref{fig2}(c) and (d)
for the large $30^o$ piles with bumpy, $L_1=97$ (c), and smooth, flat
bottom, $L_0=100$ (d). The dashed lines give the vertical stress
for the corresponding pile. In Fig.\ \ref{fig2}(c) we observe a contact
network similar to the result in Fig.\ \ref{fnet}(b) for $c=0$.
The center triangle is arranged on a diamond lattice and the
shoulders are arranged on a dense triangular lattice, i.e. the 
horizontal contacts are closed. Only close to the surface we have a
few particles on a tilted diamond lattice. In Fig.\ \ref{fig2}(d)
the situation is more complicated. We observe three regions with 
different structure. Firstly a diamond lattice in the center,
secondly a dense triangular lattice in outward direction and
thirdly, the diamond lattice tilted outwards at the ends
of the pile. In summary, we correlate the variations
of normal stress $V$ to the change of structure in the 
contact network.

\subsubsection{Variation of System Width for Smooth, Flat Bottom}

Now we vary the width of the system with flat, smooth
bottom. Here we do not vary the box width, we just increase
(or decrease) the diameters $d_1= d_L = (1+c) d_0$ of the 
left- and rightmost particles in row $M=0$. All other particles
have the fixed diameter $d_0$ so that we change the effective width 
of the system. In Fig.\ \ref{fig3} we plot the vertical
stresses $V(0), V(2), V(4)$ 
for $30^o$ piles with $L_0=22$. 
We find that the dip vanishes already for slightly increasing $c$
and relate the existence of the dip to the presence of open horizontal
bonds in the center of the pile at $X = 0.5$.
For decreasing $c$ we still observe a dip structure in the pile
but when $c$ gets too small, the stress in the array of particles 
may get asymmetric, since the pefect triangular arangement
is disturbed (see $c=-2/15$).

\PSFIGC

\subsection{Polydisperse Particles}

Starting from a monodisperse $30^o$ pile with bumpy bottom and
$L_1$ = 97, see Fig.\ \ref{fig1}(b), 
we change the particle size of each particle slowly to the
diameter $d_i = d_0 (1+r_i)$, where $r_i$ is a random number 
homogeneously distributed in 
the interval $[-r/2,r/2]$. We present the vertical stress in 
Fig.\ \ref{fdis}, for simulations with $r$ = 2/3000 (a), 2/300 (b), 
and 1/30 (c). 
We plot the result of one run (solid line) and compare it with
the monodisperse case (dashed line) and the average over 40 runs (a)
or 100 runs (b) and (c) (symbols). The fluctuations in stress
increase with increasing $r$. In fact we observe fluctuations 
much larger
than the total stress for the monodisperse pile. With increasing
$r$ the shape of the averaged vertical stress changes in the
center from a hump [see $r=2/3000$],
to a dip [see $r$ = 1/30]. The averaged stress in 
Fig.\ \ref{fdis}(c) is similar to the stress obtained (after many
averages) from a cellular automaton model for the stress propagation
in the presence of randomly opened contacts \cite{hemmingsson96}.

\PSFIGE

In Fig.\ \ref{fdis2}(a) we give the contact network of one run as
presented in Fig.\ \ref{fdis}(c). The line thickness indicates the
magnitude of forces active at a contact. In Figs.\ \ref{fdis}(b)
we present a part of this contact network, and in Figs.\ \ref{fdis}(c)  
we plot the principal axis of the stress tensor for the same part.
In Figs.\ \ref{fdis2}(a) and (b)
each line represents the normal direction of one contact and
each particle center is thus situated at the meeting point of several 
lines. Note that in Fig\ \ref{fdis2}(a) some particles inseide the 
pile have no contacts to their above neighbors, i.e. they lie below
an arch. Comparing the contact network (b) with the stresses in (c)
one may again relate the structure of the contcat network to the 
angle $\phi$ and the ratio of the principal axis, as discussed
above in Sec.\ \ref{sec:varb}.
Finally, we calculate the probability distribution $P$ for 
vertical stresses $V$. 
We average only over the lowermost row $M=1$ and also neglect
the outermost particles of this row. In detail we average 
all particles $10 \le i \le 87$ [counting from the lower
left end to the right] over 100 runs.
Since the stress in row $M=1$ is a function of $X$ in the
case of $r=0$, we scale the stresses for $r>0$ by the 
stresses found for $r=0$, i.e. we use the scaled stress
\begin{equation}
T = V^{(r>0)}/V^{(r=0)} - \min[T],
\label{eqtzz}
\end{equation}
with the minimum of all $T$, obtained from particles
which are shielded and thus feel only their own weight, see Fig.\
\ref{fdis2}.
Note that this occurs frequently, even inside the pile.
We checked that the probability distribution of $T$ does not depend 
on the specific choice of the interval, i.e. we also 
averaged over a smaller interval $33 \le i \le 67$, or 
over particles in row $M=2$ with $130 \le i \le 164$, and
we found no difference besides fluctuations. 
\PSFIGF
We plot the distribution
function $P(T)$ in Fig.\ \ref{fdis3}. The dashed line in
Fig.\ \ref{fdis3}(a) shows a power law for small stresses $T$, 
while the dotted line in
Fig.\ \ref{fdis3}(b) shows an exponential decay for large $T$.
\PSFIGG

Thus our results 
are in agreement with the theoretical predictions of Ref.\
\cite{liu95}, and the numerical findings of Ref.\ 
\cite{radjai96b}, at least for large $T$, see 
Fig.\ \ref{fdis3}(b). Note that the probabilty to find large
$T$ is greater for $r=1/30$ than for $r=2/300$, 
corresponding to stronger fluctuations. 
For small $T$ we find a power law with 
exponent -1/2 for both values of $r$, see Fig.\ \ref{fdis3}(a).

\section{Discussion and Conclusion}
\label{sec:conclusion}

We present simulations of static 2D piles made of almost
monodisperse spheres. With this simplified model we reproduce
different former theoretical predictions which were based on the 
assumption of a homogeneous contact network in the whole pile and 
perfectly rigid particles. 

One fact is that arching and the so called dip in the
vertical stress at the bottom are not neccessarily due to solid 
friction \cite{liffman94,edwards96b}. If
the contact network varies as a function of the position in the
pile we observe stresses different from the theoretical predictions
based on a regular network. If we observe arching the orientation
of the stress tensor is fixed, at least in the outer part, and
the contact network is symmetric to the center but not translation
invariant. The orientation of the major principal axis and the ratio
of the two eigenvalues of the stress tensor are correlated with
the structure of the contact network. We observe diamond lattices, 
either vertical or tilted by 60 degrees outward from the center, 
if the major principal stress is almost vertical or tilted outwards
respectively. But if the major and minor principal axis are comparable
in magnitude we observe a triangular lattice, i.e. all possible 
contacts closed, rather than a diamond lattice.
Together with the tilted contact network, i.e. strongly tilted 
principal axis, we evidence in some cases arching and a small
vertical stress under the center of the pile. If the contact
network is tilted outwards, stresses are preferentially 
propagated outwards, what may be regarded as a reason for arching 
and for the dip.

Varying the size of the particles randomly, we find that 
already tiny polydispersities destroy the regular contact network.
Due to the small fluctuations in particle size the particles
are still positioned on a triangular lattice even when the contacts
are randomly open. In the case of a random network we also
find the so called stress chains, i.e. selected paths of
large stresses, and the stress fluctuations are larger or of the order
of the mean stress. The stress chains - or better the stress network 
- is also disordered.
When averaging over many realizations of the stress network
we get a dip in the vertical stress at the bottom if the size 
fluctuations are large enough. 
Thus we observe a similar stress distribution at the bottom
as obtained by a cellular automaton model based on a random opening
of contacts.

Since we are able to find most of the phenomenology, expected in a 
sandpile, already in an oversimplified regular model system, we
conclude that the role of the contact network (or the fabric)
is eminent. However, friction and small polydispersity may play
a different role in more general situations with physical sandpiles.

As an extention of our model we started more 
realistic simulations with a nonlinear Hertz contact law
\cite{luding97b}, with
solid friction, and also with nonspherical particles
\cite{matuttis97}. The effect
of those more realistic interaction laws has to be elaborated and
also threedimensional examinations should be performed.

\section*{Acknowledgements}

We thank J. D. Goddard, H.-G. Matuttis, H. J. Herrmann, and 
J. J. Wittmer for helpful discussions,
and acknowledge the support of the European network ``Human Capital
and Mobility''  and of the DFG, SFB 382 (A6).


\end{document}